%% file: paper.tex
\documentclass[12pt,preprint]{aastex}

\shorttitle{Three Edge-on Massive YSOs}
\shortauthors{Simpson et al.}

\begin{document}

\title{{\it Hubble Space Telescope} NICMOS Polarization Observations of Three Edge-on 
Massive YSOs\footnote{Based on observations made with the NASA/ESA Hubble Space Telescope, 
obtained at the Space Telescope Science Institute, which is operated by the Association 
of Universities for Research in Astronomy, Inc., under NASA contract NAS 5-26555.
These observations are associated with program \#10519.}
}

\author{Janet  P. Simpson\altaffilmark{1,2}, Michael G. Burton\altaffilmark{3}, 
Sean W. J. Colgan\altaffilmark{1}, Angela S. Cotera\altaffilmark{2}, 
Edwin F. Erickson\altaffilmark{1}, Dean C. Hines\altaffilmark{4}, 
and Barbara A. Whitney\altaffilmark{4}}

\altaffiltext{1}{NASA Ames Research Center}
\altaffiltext{2}{SETI Institute}
\altaffiltext{3}{University of New South Wales}
\altaffiltext{4}{Space Science Institute}

\email{janet.p.simpson@nasa.gov}

\begin{abstract}

Massive young stellar objects (YSOs), like low-mass YSOs, 
appear to be surrounded by optically thick envelopes and/or disks and
have regions, often bipolar, 
that are seen in polarized scattered light at near-infrared wavelengths.
We are using the $0.2''$ spatial resolution of NICMOS on {\it Hubble Space Telescope} 
to examine the structure of the disks and outflow regions of massive YSOs 
in star-forming regions within a few kpc of the Sun. 
Here we report on 2 \micron\ polarimetry of NGC 6334 V and S255 IRS1.
NGC 6334 V consists of a double-lobed bright reflection nebula seen against 
a dark region, probably an optically thick molecular cloud.
Our polarization measurements show that the illuminating star lies $\sim 2''$
south of the line connecting the two lobes;
we do not detect this star at 2 \micron, 
but there are a small radio source and a mid-infrared source at this location.
S255 IRS1 consists of two YSOs (NIRS1 and NIRS3) 
with overlapping scattered light lobes 
and luminosities corresponding to early B stars. 
Included in IRS1 is a cluster of stars 
from whose polarization we determine the local magnetic field direction. 
Neither YSO has its scattered light lobes aligned with this magnetic field.
The line connecting the scattered light lobes of NIRS1 is twisted symmetrically 
around the star; the best explanation is that the star is part of a close binary 
and the outflow axis of NIRS1 is precessing as a result of non-coplanar disk and orbit.
The star NIRS3 is also offset from the line connecting its two scattered light lobes. 
We suggest that all three YSOs show evidence of episodic ejection of 
material as they accrete from dense, optically thick envelopes.
\end{abstract}


\section{Introduction}

Although there has been substantial progress in understanding the formation of 
solar mass stars in recent years (see, e.g., Klein et al. 2007 for a review),
the formation of massive stars is still not understood.
Low-mass stars almost certainly form by accretion through a disk onto a dense core, 
or protostar;
these young stellar objects (YSOs) are observed as disks surrounding the more or less visible 
protostar with an outflow or sharp jet emanating along the disk axis (e.g., Burrows et al. 1996).
The theoretical difficulty regarding massive star formation 
is that hydrogen fusion commences before the star reaches the main sequence 
once the accreting YSO gets to about 8 -- 10 M$_\odot$ 
and the resulting high luminosity should stop the accretion process 
(Kahn 1974; Stahler et al. 2000);
recent work, however, on accretion through disks suggests that 
the massive YSO can continue accreting beyond 8 M$_\odot$ if 
it also has outflows that produce holes through which the 
radiation can escape (Krumholz et al. 2005).
The holes occur when the accretion disk changes the direction of radiation output 
from isotropic to concentrated along the disk axis 
(this has been called the ``flashlight effect'', Yorke \& Sonnhalter 2002).
This theory is in accordance with the observations of 
a number of objects with luminosities $10^4 < L/L_\odot < 10^5$ 
that are described as massive YSOs (see, e.g., Table~5 of Zinnecker \& Yorke 2007).
These luminosities correspond to early B stars;
the observed YSOs appear to have substantial outflows emanating from 
still deeply embedded stars or protostars. 
However, since such disk structures with outflows 
have not yet been detected for objects with $L > 10^5 L_\odot$, 
Cesaroni et al. (2006) suggested that some other star formation process 
might be required, such as mergers of lower mass stars 
(e.g., Bonnell \& Bate 2002; Bally \& Zinnecker 2005) 
or accretion through the extra-deep potential well of a massive star cluster 
(Bonnell et al. 2004; Bonnell \& Bate 2006). 

Since it is essential to understand 
the relation between massive YSOs' outflows and disks,
we have undertaken a study of such systems with {\it Hubble Space Telescope} ({\it HST}), 
using the near-infrared (NIR) polarimetry capability of 
its Near Infrared Camera and Multi-Object Spectrometer (NICMOS).
For this study we chose the YSOs and candidate YSOs closest to the Earth 
whose luminosities indicate that their masses are $> 8$ M$_\odot$
(log $L/L_\odot \gtrsim 3.5$). 
(Unfortunately, there are no NIR-visible, $L > 10^5$ L$_\odot$ YSOs 
close enough that a $\sim 500$ AU disk could be resolved with the $0.2''$ resolution
of NICMOS at 2~\micron.) 
In this paper we report on two locations, NGC 6334 V and S255 IRS1, 
which have three massive YSOs whose disk axes are close to the plane of the sky. 
Our goals are to use polarimetry to pinpoint the location of the illuminating star, 
characterize the structure of any circumstellar disks and outflow regions, 
and determine the orientation of the local magnetic field 
in the plane of the sky through measurements of the polarization 
of the other stars in the field of view.

NGC 6334 V is a luminous far-infrared source 
(source number V of McBreen et al. 1979) at a distance of $\sim 1.74 \pm 0.31$ kpc (Neckel 1978).
See the Appendix of Kraemer et al. (1999) for the nomenclature of the sources in NGC 6334.
Measurements of its total luminosity range from 
$6.5 \times 10^4$  L$_\odot$ (Harvey \& Gatley 1983) to $1.7 \times 10^5$ L$_\odot$ 
(Loughran et al. 1986) and $1.9 \times 10^5$ L$_\odot$ (McBreen et al. 1979), 
depending on the effective aperture size of the measurement
(the larger luminosities probably include multiple sources). 
It is coincident with a compact molecular cloud seen in CO, CS and NH$_3$ 
(Plume et al. 1992; Jackson \& Kraemer 1999; Kraemer \& Jackson 1999).
Harvey \& Wilking (1984) showed that there is a NIR bipolar source at this location
projected against a dark background, undoubtedly the optically thick molecular cloud.
With higher resolution Simon et al. (1985) showed that the brighter, eastern lobe 
has at least three NIR components, which they named NGC 6334 V-1, V-2, and V-3 
ordered from east to west. 
Chrysostomou et al. (1994) and Nakagawa et al. (1990) found that 
the bipolar source is highly polarized, indicating that it is a reflection nebula.
The illuminating star itself has not been detected at NIR wavelengths, 
although Simon et al. (1985) and Chrysostomou et al. (1994) 
suggest that the westernmost condensation in the eastern bipolar lobe, NGC 6334 V-3, 
could be a candidate. 
The appearance is quite different at mid-infrared (MIR) wavelengths, 
where four compact MIR sources are observed (Kraemer et al. 1999). 
At 15 GHz three very compact radio sources are seen in the vicinity, 
but none of the radio sources has the appropriate flux for an early-type main sequence star 
(Rengarajan \& Ho 1996; Jackson \& Kraemer 1999).
There is no one-to-one correspondence between the NIR, MIR, and radio sources.
In particular, NGC 6334 V-3 has no associated radio source. 
Other indications of the presence of an unidentified massive star  
include photodissociation-region emission (H$_2$ and PAHs) in the surrounding area 
(Burton et al. 2000; Rathborne 2003).

S255 IRS1 is the most luminous source ($\gtrsim 2.6 \times 10^4$ L$_\odot$, Jaffe et al. 1984) 
in the star-forming region located between the S255 and S257 H II regions
(Beichman et al. 1979).
IRS1 is located in the center of a cluster of young stars and YSOs, 
which is seen at NIR wavelengths (Howard et al. 1997), 
and includes at least three lobes of diffuse polarized light on either side 
of the two brightest stars 
(Tamura et al. 1991; Itoh et al. 2001).
These two YSOs, NIRS1 and NIRS3, are separated by about $2.5''$ 
(6000 AU at a distance of 2.4 kpc, Chavarr\'{i}a et al. 2008),
both have very red spectral energy distributions (SEDs) 
indicative of high extinction probably caused by disks or thick envelopes, 
and both have reflection nebulosity quite close to the star (Alvarez et al. 2004).
NIRS3, the more luminous YSO, also emits photons able to ionize the local gas 
(ionizing photons have energy $> 13.6$ eV).
This is seen both in the radio (Rengarajan \& Ho 1996; Snell \& Bally 1986) 
and in Br$\alpha$ and Br$\gamma$ (Howard et al. 1997).
NIRS3's scattering lobes are also visible in Br$\gamma$ (Howard et al. 1997).
There are knots of H$_2$ $\nu = 1 - 0$ S(1) emission 
aligned with the outflow from NIRS3 in both the east and west directions, 
as well as strong blue-shifted CO emission at the location of NIRS1 and NIRS3 
and weaker blue-shifted CO emission located approximately at the west H$_2$ knots
(Miralles et al. 1997).
Additional, younger or more deeply embedded star forming regions 
occur about $13''$ east and $1'$ north of IRS1 
and are known as S255 IRS2 or G192.60:2-3 (Beichman et al. 1979; Longmore et al. 2006) 
and S255-N (Jaffe et al. 1984), respectively.
Snell \& Bally (1986) name the co-located northern radio source S255-1,
and the sources associated with S255 IRS1 and IRS2 they call S255-2. 

In this paper we present  2.0 \micron\ polarimetric imaging of NGC 6334 V and S255 IRS1 
with NICMOS. 
We describe the observations in \S2, including those of the very red standard star Oph-N9. 
In \S3 we discuss the results for both NGC 6334 V and S255 IRS1, 
and in \S4 we summarize our conclusions.

\section{Observations}

NGC 6334 V and S255 IRS1 were observed for two visits each with NICMOS on {\it HST} 
with the Camera 2 POL0L, POL120L, and POL240L filters 
(hereafter the ``POL filters''), which cover a 1.9 -- 2.1 \micron\ bandpass
($0.2''$ resolution).
For the purpose of point-spread function (PSF) subtraction, 
we also observed the red standard star Oph-N9
(Ks = 9.620, H$-$K = 2.862, Persson et al. 1998).
All objects were observed with a spiral dither pattern with spacing $1.0213''$, 
with NGC 6334 V and S255 IRS1 having six dither positions and Oph-N9 having four.
A journal of the observations is given in Table~1.
Oph-N9 is also known as GY232 (Greene \& Young 1992) and
BKLT 162713$-$244133 (Barsony et al. 1997) and has been spectroscopically
observed to be a background giant (Luhman \& Rieke 1999).
We measure the polarization, $P$, to be 8.5\% with polarization position angle $\theta = 34^\circ$ ---
this polarization is due to its extinction by the $\rho$ Oph cloud.

The detector array was read out in MULTIACCUM mode
with sample sequence STEP16 to accumulate total times of 95 or 112 s per integration
(48 s for Oph-N9),
with the POL240L measurements having less integration time because this filter has better 
transmission.
Each image was reduced (dark subtracted, flat fielded, etc.)
with Space Telescope Science Data Analysis System (STSDAS) task CALNICA
(equivalent to the {\it HST} pipeline).
However, each image has bright pixels that produce strong ghost images with accompanying
vertical streaks due to amplifier ringing (also known as the ``Mr. Staypuft'' anomaly,
McLaughlin \& Wiklind 2007).
These electronic ghosts can be and largely were removed by performing 
the STSDAS task PUFTCORR on the raw data prior to running CALNICA.

The NICMOS images need additional corrections for bad pixels, the {\it HST}
thermal background, and ``Pedestal'', which is a noiseless bias that can
vary from quadrant to quadrant and readout to readout.
The bad pixels (about 50 plus the center column and much of the bottom row)
were removed by substituting the median of surrounding pixels before shifting.
The {\it HST} thermal background was removed by subtracting the nominal
background computed by the NICMOS Exposure Time Calculator
($\lesssim 0.2$ DN s$^{-1}$) and then adding back in a small constant if
subtracting the nominal thermal background produced a region of negative
flux in the image.
The minimum flux was computed for each image by measuring the median
of $\sim 800$ pixels centered on exactly the same position on the sky.
Since these minima are never the same, small constants ($\sim 0.01 - 0.02$ DN s$^{-1}$)
were subtracted from those images with the larger minimum fluxes
so that all images would have the same flux levels for later median combining.
These variations probably are due to Pedestal contributions. 
The ``HST Data Handbook for NICMOS'' (McLaughlin \& Wiklind 2007) 
describes several STSDAS tasks that can be used to correct for Pedestal, 
but they produce spurious results for images 
that have as much structure in the diffuse emission as ours. 
Four of the 72 images showed noticeable jumps from quadrant to quadrant 
that can also be ascribed to Pedestal. 
These jumps were removed by subtracting constants (0.03 -- 0.05 DN s$^{-1}$)
from the affected quadrants (quadrant to quadrant jumps $\sim 0.01$ DN s$^{-1}$
are present in all images but are difficult to determine owing to the
nebulosity, are generally not noticeable, and so are ignored).
The result of this analysis is an uncertainty $\lesssim 0.03$ DN s$^{-1}$
in the continuum level for each image. 
However, a value of 0.03 DN s$^{-1}$ in Stokes $I$ 
corresponds to only 0.016 mJy arcsec$^{-2}$; 
this is substantially fainter than any of the nebulosity that we are measuring. 
We obviously cannot measure the polarization of the background. 

For each polarizing filter, we aligned and shifted the six dither positions
by centroiding two bright stars
using the IDL program, 
IDP3\footnote{http://mips.as.arizona.edu/MIPS/IDP3/idp3setup.html}.
The shifted images (with very small additions/subtractions to the flux that
represent variations in the ``Pedestal'', McLaughlin \& Wiklind 2007,  
or {\it HST}  thermal corrections)
were then median-combined to remove the remaining bad pixels;
the result is three images for the three POL filters in each visit, 
all aligned to the same position on the sky.
Because the original plate scale was $0.075948''$ by $0.075355''$ per
pixel, the pixels were rectified to $0.07595''$ per pixel. The images
were also smoothed with a $3\times3$ box-car (i.e., to approximately the
{\it HST} spatial resolution of $0.2''$ at 2.0 \micron) in the data
reduction process to achieve higher signal/noise for $P$ and $\theta$
images. However, the intensity images plotted in this paper are not smoothed
in order to expose details such as the fine structure of the {\it HST}
diffraction pattern; the plotted polarization images and the over-plotted
polarization position angle vectors are from the smoothed data.

After rectification and smoothing, Stokes $I$, $Q$, and $U$ intensities were
computed from the reduced data (Hines et al. 2000; Batcheldor et al. 2006),
and rotated so that north is up. The resulting celestially-aligned Stokes
$I$, $Q$, and $U$ images from each pair of visits were then mosaicked
together and the fractional polarization $P$ and the position angle $\theta$
of the polarization vectors were computed from the mosaicked $I$, $Q$, and
$U$ using the usual relationships $P = (Q^2+U^2)^{0.5}/I$ and 
$\theta = 0.5\ {\rm arctan} (U/Q)$.
These mosaics, with polarization vectors and fractional polarization, 
are shown in Figures 1 and 2.

The positions, polarization parameters, and fluxes were measured for every detected 
star in the images. 
These values are given in Tables 2 and 3 and the locations of the stars 
are plotted in Figures 1c and 2c.
The positions of the stars were measured by fitting Gaussian functions to the 
cores of the star images in the unsmoothed mosaics (Figs. 1b and 2b).
Relative positions for all but the faintest stars are accurate to $\sim 0.02''$.
Absolute positions are much less accurate and 
will be discussed in the section applying to each cluster.
The polarization parameters were measured by aperture photometry of each star on 
the median-combined but unshifted and unrotated POL images; 
$I$, $Q$, $U$, $P$, and $\theta$ were computed from the measured fluxes 
by multiplying the flux vector by the same matrix that was used to 
compute $I$, $Q$, and $U$ from the combined dithered images. 
The circular aperture for the star measurement had a radius of 2.5 pixels (to the 
minimum of the Airy dark ring) and the background was measured in 
a ring of radii 5 to 7 pixels (just outside the first Airy bright ring).
Batcheldor et al. (2006) plot the accuracy of measurements of $P$ and $\theta$ 
as a function of the radius used in aperture photometry for both simulated 
star images with sub-pixel misalignment and for isolated polarization standard stars. 
From their plots we estimate that not using a larger photometry aperture introduces 
an uncertainty of a few tenths of a percent to the measured percentage polarization; 
however, we feel that the smallest aperture possible should be used due to the 
large and variable background in the images.
Moreover, since the difference between the visits is often larger than a percent
(in which case the error in the tables is equal to 0.5 times the difference 
of the measured values), we express the results as percent without a decimal point.
The fluxes were taken from aperture photometry on the computed Stokes $I$ images and 
multiplied by correction factors 
to account for our small aperture compared to the aperture used in the NICMOS calibration.
See Simpson et al. (2006) for further description.

We also compute uncertainties for the stellar polarization measurements 
from first principles. 
We start by estimating the uncertainties in $Q$ and $U$, $\sigma_Q$ and $\sigma_U$, 
in a single pixel as a function of flux, given as photon counts per exposure,
using the average signal in the photometry apertures for the source 
and background fluxes per pixel.
These uncertainties we take from Figure 10 of Hines et al. (2000), who  
simulated the uncertainties in $\sigma_Q$ and $\sigma_U$ 
as a function of photon flux through Monte Carlo calculations
that included the non-ideal parameters of the NICMOS POL filters
(different transmissions for each filter, imperfect filter orientation),
read noise, photon noise, and the calibration uncertainties.
These $\sigma_Q$ and $\sigma_U$ are then reduced by the square root of 
the number of pixels in the star and background apertures, 
the number of dithers (6), and the number of visits (2).
Using the quadratic sum of the source and background uncertainties, 
we compute the uncertainties in $P$ and $\theta$, $\sigma_P$ and $\sigma_\theta$, 
using the equations derived by Sparks \& Axon (1999), who 
analyzed systems of three polarizer filters, such as are found in NICMOS.
In general, the uncertainty in $P$ is inversely proportional to $I$,
but the uncertainty in $\theta$ is inversely proportional to $IP$ ($I$ times $P$).
Typical results are that an 18.2 magnitude star with $P = 10$\% 
has $\sigma_P = 2.3$\% and $\sigma_\theta = 8^\circ$
and a 14.7 magnitude star with $P = 4.5$\% 
has $\sigma_P = 0.2$\% and $\sigma_\theta = 1.2^\circ$.
The systematic uncertainties (e.g., difference between visits) 
are usually larger than this.

The polarizations and uncertainties are given in Tables 2 and 3.
Values for polarization are listed only for stars with $P/\sigma_P > 5$ 
with the additional requirement that the star be detected on the  
$Q$ and $U$ images.
Because of the possibility of systematic uncertainties, the minimum 
tabulated $\sigma_P$ is 1\%; however, the stars in the tables 
with $P = 1 \pm 1$ and $P = 2 \pm 1$ are real detections.
These are relatively bright stars and the computed uncertainties in $P$ 
are $\lesssim 0.1 - 0.3$\%. 
The uncertainties for the fluxes should include an uncertainty of order 5\% -- 10\%  
due to uncertainties in the calibration of the POL filters 
(see the discussion of the NICMOS flux calibration uncertainties at
 http://www.stsci.edu/hst/nicmos/performance/photometry).

\section{Results and Discussion}

\subsection{NGC 6334 V}

Figure 1 shows two highly polarized reflection nebulae 
connected by a faint ``bridge'' of polarized nebulosity.
These polarized reflection nebulae have previously been discussed by 
Simon et al. (1985), Nakagawa et al. (1990), Chrysostomou et al. (1994), 
and Hashimoto et al. (2007).
However, our NICMOS data have either much higher spatial resolution 
or much higher sensitivity --- 
none of the previous observations detects the emission, let alone the polarization, 
in the bridge, and we can detect stars at least two magnitudes 
fainter than any previous detections. 
Moreover, 
nowhere within the two reflection nebulae do we detect any point sources 
(sources exhibiting the {\it HST} diffraction pattern) 
that could be investigated for identification as the star that is illuminating  
the nebulae.

In the past, the lack of absolute astrometry for NIR and MIR images 
has resulted in authors' claiming coincidence of two blobs of nebulosity 
detected at widely differing wavelengths 
and from this claim of coincidence, adjusting the coordinates of the rest of the image 
to identify other ``stars'' as the illuminating star for the observed reflection nebulae
(e.g., Kraemer et al. 1999; Hashimoto et al. 2007).
We can do a little better here --- 
thanks to the diffraction limited resolution of NICMOS on {\it HST}, 
we are able to identify what sources are truly stars and 
what are blobs of nebulosity, whether the self-illuminated diffuse nebulosity of a YSO
(e.g., the YSO labeled as such in Fig.~2)  
or a dense clump of dust illuminated by an exterior star.

In this paper we locate sources by offsetting from objects 
whose absolute coordinates we deem the most reliable. 
For the NICMOS images of NGC 6334 V (Table 2) this is a star from the Two Micron All Sky Survey 
(2MASS, Skrutskie et al. 2006), 2MASS17195766$-$3557416,
which has a tabulated uncertainty of $\leq 0.07''$.
For radio sources we use the VLA measurements of Rengarajan \& Ho (1996), 
which have an uncertainty of $0.1''$.
For MIR positions we use the Spitzer IRAC images from the GLIMPSE Legacy survey
(Benjamin et al. 2003), 
which have an uncertainty of $< 0.2''$ (GLIMPSE astrometry is derived from 2MASS).

Using 2MASS17195766$-$3557416, we have estimated the absolute coordinates of Figure 1.
In this figure and also in Figure 3a we have plotted  
the positions of the radio sources measured by Rengarajan \& Ho (1996).
The coordinates and average polarization of the clumps observed by Simon et al. (1985) 
are also given in Table 2.
We add a fourth bright clump, NGC6334 V-4. 
Note these are not the clumps of similar names of Hashimoto et al. (2008).
There are {\it no} stars visible within the reflection nebulosity ---  
the previous candidates for the illuminating star V-3 
(Simon et al. 1985; Chrysostomou et al. 1994) 
and WN-A1 (Hashimoto et al. 2007) are too extended to be stellar. 

We have computed the location of the illuminating star from the intersection 
of the perpendiculars to the polarization vectors given in Figure~1:  
$17^{\rm h}19^{\rm m}57\fs38$ $-35^\circ 57' 51.9''$, plotted as the diamond 
in Figures~1b, 1c, and 3a.
The position of the star that is illuminating the two lobes of reflection nebulosity 
is located within the errors at the radio source R-E3 of Rengarajan \& Ho (1996).
Although this star cannot be detected at 2 \micron, 
it is visible at longer wavelengths. 
This is shown in Figure 4, which is an IRAC image from the 
Spitzer GLIMPSE Legacy survey (Benjamin et al. 2003).
The two bipolar lobes are colored in blue and green  against an 
extended dark, starless region that must correspond to the dense molecular cloud.
Although the 3.6 \micron\ emission is almost certainly scattered light 
(the shorter wavelengths scatter light better), 
the 4.5 \micron\ emission could also arise from a shocked outflow (CO or H$_2$),
similar to that seen in several young and massive protostars 
in the GLIMPSE survey (Cyganowski et al. 2008).
At the longest wavelengths (5.8 and 8.0 \micron) a red source appears 
at the same RA and Dec as the illuminating star. 
This source, labeled MIR Source in Figure 4, is probably also the 12.5 \micron\ source 
KDJ4 detected by Kraemer et al. (1999) (the absolute coordinates of KDJ4 are not well determined, 
K. Kraemer 2006, private communication).
The MIR source, however, is not readily apparent at 20 \micron\ (Simon et al. 1985),
from which we infer that its actual temperature is much hotter than typical dust temperatures.

\subsubsection{Dust Grain Parameters and Models}

Only the perpendiculars from the regions with the highest polarization (Fig.~1) 
point directly to the illuminating star --- 
these regions are where the polarization is due to single scattering. 
Regions where there is multiple scattering have lower polarization 
and the polarization position angle is more indicative of the 
location of the last scattering than of the original illuminating star.
This is seen in 
polarization models of YSOs with optically thick envelopes and outflows, 
which generally show that 
the outflow regions produce substantial scattered, low optical depth light 
with polarization position angles perpendicular to the direction to the 
illuminating star/YSO (e.g., Fischer et al. 1996; Bastien \& M\'enard 1988; 
Whitney \& Hartmann 1993; Whitney et al. 1997).

Optical depth effects are important when 
the outflow axis is close to the plane of the sky such that 
the rotationally flattened envelope is close to edge-on. 
Both the amount of light and the polarization are reduced 
along the center line that is perpendicular to the outflow axis
because of the high optical depth in the envelope.
Moreover, the polarization position angles are parallel to this center line
and are not perpendicular to the direction to the illuminating star
because the light that is scattered towards the observer comes not from the star
but from locations well above or below the flat, disk-like envelope
(that is, it is multiply scattered).
This configuration of polarization position angles has been called a 
``polarization disk'' (Whitney \& Hartmann 1993).
However, we prefer to call it a ``parallel polarization pattern'' 
(Wolf et al. 2002) to emphasize the fact 
that an actual, physical ``disk'', which Cesaroni et al. (2007) define as
``a long-lived, flat, rotating structure in centrifugal equilibrium,''
is not required.
Note that recent work 
(e.g., Allen et al. 2003; Hennebelle \& Fromang 2008; Mellon \& Li 2008) 
has shown that the presence of even a weak magnetic field 
inhibits the formation of a thin, rotationally supported ``disk'' 
owing to magnetic braking, leaving only 
a ``pseudodisk'' (Galli \& Shu 1993).

We show an example of an edge-on (inclination of $87^\circ$), optically thick model 
with an outflow 
produced using the Monte Carlo dust scattering code of Whitney \& Hartmann (1993) in Figure 3b.
In order to show that a physical disk is not necessary to produce a parallel polarization 
pattern, this model has  {\it no} disk (a model with a disk looks very similar).
This model was produced for a YSO with mass 10 M$_\odot$, 
accretion rate $2\times10^{-4}$ M$_\odot$ yr$^{-1}$.
We consider the situation where the extinction in front of NGC 6334 V is asymmetric, 
such as might be due to the dense cloud seen in NH$_3$ by Kraemer et al. (1999).
(This NH$_3$ cloud could, in fact, be the densest part of an asymmetric YSO envelope.)
In Figure 3b this cloud is plotted as a gray disk
to indicate the area that would be hidden by the additional extinction.
The model with the asymmetric extinction has a remarkable resemblance to 
the appearance of NGC 6334 V in Figure 3a --- 
the apparent ``bridge'' between the lobes is almost certainly due 
to lower extinction.

The two lobes must lie essentially in the plane of the sky 
because the maximum polarization of the eastern lobe is so high --- 95\%.
The maximum polarization for light scattered at any given angle 
is a function of the shape of the particle 
(scattering by non-spherical, aligned grains 
can produce higher polarization than scattering by spherical grains 
of the same size and composition, Whitney \& Wolff 2002)
and the size (radius) of the particle, such that for particles much smaller 
than the wavelength (about an order of magnitude), 
the maximum polarization at scattering angles near $90^\circ$ is almost 100\%.
To compare with observations, the scattering cross sections must be integrated 
over the particle size distribution.
Thus the fact that we detect polarization as high as 95\% means that 
(1) the scattering angle is close to $90^\circ$, 
(2) the light has been scattered only once 
(multiple scattering decreases the polarization), and 
(3) the particle size distribution is dominated by small particles, as it is in 
a normal interstellar size distribution (e.g., Mathis et al. 1977, hereafter MRN). 

Envelopes of YSOs, such as in Figure 3b, 
are usually modelled with the grain size distributions of dense clouds, 
since in dense clouds the relative number of large particles is increased, 
probably by coagulation in the dense regions of the envelope. 
A particle size distribution with larger particles 
has a larger extinction ratio $R_V = A_V/E_{B-V}$ 
(Kim et al. 1994; Weingartner \& Draine 2001) 
and a smaller maximum polarization due to scattering  
(see, for example, Fig.~2 of Fischer et al. 1994 and Fig.~13 of Lucas \& Roche 1998).
However, our observation of $\sim 95$\% polarization is not compatible 
with this size distribution. 
We suggest that a possible explanation for the need for much smaller grains than are found 
in dense clouds is that the larger grains are destroyed in fast shocks in the outflows.
Moreover, the dust in the envelope must be clumpy with holes so that more 
photons can get out after only one scattering. 

The model in Figure 3b used a particle size distribution
similar to that of Kim et al. (1994) for $R_V = 4.0$, 
which distribution includes significant numbers of grains with sizes approaching 1 \micron.
Because of the large particles, the maximum polarization of the model is 
only $\sim 42$\%, much smaller than is observed.
We computed a number of other models to test the effects of grain size distribution
on maximum polarization and production of a parallel polarization pattern.
No significant parallel polarization pattern is produced 
for a diskless model computed with a MRN size distribution 
(which is dominated by small particles) instead of a Kim et al. distribution. 
The reason is that the small MRN particles, 
while producing a higher polarization, have a lower albedo.  
The parallel polarization vectors in the disk midplane in Figure 3b 
are caused by multiple scattering: the disk midplane is 
too optically thick for radiation to
propagate directly out and instead diffuse emission in the midplane has scattered at least
once in the less opaque bipolar cavities, 
before its final scattering in the outer regions of the disk midplane.   
The MRN grains produce little multiple scattering due to their lower albedo,
and therefore have little or no emission arising in the disk midplane, 
and produce no parallel polarization pattern.

In order to reproduce both the high polarization and the parallel polarization pattern, 
the models appear to require both small (high intrinsic polarization) 
and large (high albedo) grains.  
We produced the models described above using the scattering parameters of spherical grains. 
Elongated grains produce higher polarization than spherical grains, even
when the size distribution function includes the larger grains of Kim et al. (1994) 
(Whitney \& Wolff 2002; Wolf et al. 2002). 
We next consider the appearance of a YSO envelope containing elongated grains, 
which grains are aligned by the local magnetic field.

For low mass stars, Shu et al. (1987) proposed that 
the magnetic field is parallel to the outflow axis.
An axial magnetic field in the disk midplane would produce polarization vectors 
parallel to the outflow axis (Whitney \& Wolff 2002), which is perpendicular to 
the vectors of a parallel polarization pattern.
This pattern is not commonly observed, even in low-mass YSOs (e.g., Lucas \& Roche 1998).
Whitney et al. (2009, in preparation) have computed models with magnetically aligned, 
elongated grains 
that include a toroidal field in the disk midplane and an axial field in the upper cavities; 
these models have polarization patterns that resemble a polarization disk 
perpendicular to the outflow axis and twisted polarization vectors in the outflow region. 
Since there is confusion of polarization caused by scattering and 
polarization caused by dichroic absorption at NIR wavelengths, 
the magnetic field directions that are the input to these models could be tested by 
high spatial resolution observations of the MIR polarization (e.g., Aitken et al. 1993) 
or the far-infrared/sub-mm polarization (e.g., Curran \& Chrysostomou 2007) 
of massive YSOs.
Such tests will help characterize the grain properties in YSO envelopes and outflows. 

\subsubsection{Possible Multiple Stellar Components}

To summarize the geometry of NGC 6334 V, 
we find that the illuminating star of the bipolar lobes 
seen at NIR wavelengths has the same coordinates, within the uncertainties, 
as the radio source R-E3 
and the MIR source (KDJ4) seen in the IRAC 8 \micron\ image (Fig. 4).
We conclude that these are all the same object. 
The scattered light lobes almost certainly coincide with the outflows from this YSO, 
since there is emission from shocked H$_2$ at 2.12 \micron\ (Rathborne 2003)
at the same locations.
It is possible that the lobe emission seen at 4.5 \micron\ in Figure 3 is also due to 
shocked H$_2$ (Marston et al. 2004; Noriega-Crespo et al. 2005; 
Smith et al. 2006; Cyganowski et al. 2008).
The fact that the illuminating star is  $\sim 2''$ south of the centerline of the lobes
is almost certainly due to extinction.
We conclude that the scattered light lobes, and hence the outflow axis, 
are all at inclinations close to $90^\circ$ to the line of sight. 

However, there are indications of  
a second, less evolved YSO in NGC 6334 V located a few arcseconds north 
of the 8 \micron\ source and the 2 \micron\ illuminating star.
This YSO would produce the outflow, seen in CO, 
that is close to parallel to the line of sight, 
with the blue-shifted CO line wing slightly north and the red-shifted CO line wing 
slightly south of the outflow center (Kraemer \& Jackson 1995, 1999).
We position it further north because 
Harvey \& Gatley (1983), Harvey \& Wilking (1984), Simon et al. (1985), and Kraemer et al. (1999)
consistently plot the peak of the intensity at 20 \micron\ 
several arcseconds north of radio source R-E3, the 8 \micron\ IRAC source, and KDJ4
(all the same object) in their figures. 
Moreover, the 8 \micron\ IRAC image has a faint extension several arcseconds north-east 
of the main MIR source (identified with an arrow in Fig. 4 as a possible second YSO).
For absolute coordinates, observed with a $19''$ beam, 
Midcourse Space Experiment source 
(MSX, Egan et al. 2003)
MSXC6\_G351.16124+00.6973, which is brightest at the longest MSX wavelength (21 \micron),
has coordinates 17$^{\rm h}$19$^{\rm m}$57\fs48$-35^\circ 57' 49.3''$,
or $2''$ north of the illuminating star ($3''$ uncertainty).
More significantly, the centroid of this source (measured on MSX images, which have $6''$ pixels)
moves north by about $1''$ going from  the Band A (8 \micron) to the Band E (21 \micron) images.
This source, IRAS 17165-3554, has its maximum measured flux between 50 and 100 \micron\ 
(Harvey \& Gatley 1983), characteristic of a very young YSO.
Finally, although the peak of the NH$_3$ source measured by Kraemer \& Jackson (1995) 
can be associated with KDJ4 and radio source R-E3, 
there is a secondary NH$_3$ peak $\sim 2''$ to the northeast which can be associated with 
KDJ3, the brightest source in NGC 6334 V (Kraemer et al. 1999). 
We suggest that KDJ3 is associated with this second YSO
(see also the discussion of Kraemer et al. 1999). 

In this interpretation, the thick envelope/disk of the less-evolved YSO could be 
a major contributor to the asymmetric extinction that seems to be required 
to explain the 2 \micron\ appearance of and illuminating-star location of 
the more evolved YSO.
There are additional reasons for suspecting the presence of a second YSO.
The total luminosity of NGC 6334 V is $\sim 6.5 - 17 \times 10^4$L$_\odot$, 
which corresponds to a late O star, if on the main sequence (see, e.g., Martins et al. 2005). 
However, it certainly is not on the main sequence since the number of ionizing photons, 
estimated to be $\sim 3 \times 10^{45}$s$^{-1}$ (using equation 1 of 
Simpson \& Rubin 1990), corresponds to approximately a B2 star.
The discrepancy between the ionizing luminosity and the total luminosity 
(as was remarked upon by Rengarajan \& Ho 1996) could be due to the presence 
of more than one star.

We refer to the two stellar components as the more evolved YSO and the less evolved YSO.
It is the more evolved YSO (the illuminating star) that
is close enough to the main sequence that it is 
producing the energetic photons that ionize R-E3, and possibly the other 
radio sources (R-E1 and R-E2, Rengarajan \& Ho 1996) found in the east lobe.
The NIR sources V-1, V-2, and V-3 of Simon et al. (1985) are clumps of dust
in the outflow that are heated and scatter light at 2 \micron.
Since dust clump V-1 lies very close to radio sources R-E1 and R-E2 (Fig. 3a),
the gas producing these radio sources may be ionized by fast shocks 
(e.g., McKee \& Hollenbach 1987) instead of photoionized by the more-evolved YSO.
However, the bulk of the luminosity would come from a less-evolved 20 \micron\ and 
far-infrared source. 
The luminosity of either source is probably not a good indicator of its mass, since 
most of the luminosity of such objects is the accretion luminosity 
and not the luminosity from nuclear fusion of a star on the main sequence 
(see, e.g., the review by Zinnecker \& Yorke 2007).
We defer further discussion of these interesting topics to a later paper.
 
\subsection{S255 IRS1}

Figure 2 shows the intensity and fractional polarization images of S255 IRS1.
This image is dominated by the scattered light lobes of the two brightest YSOs, 
NIRS1 and NIRS3, plus a number of stars of their surrounding stellar cluster.
The scattered light lobes illuminated by NIRS1 extend north and south of the YSO, 
and the scattered light lobes illuminated by NIRS3 extend to the east 
at position angle $\sim 67^\circ$ and to the southwest at 
position angle $\sim 231^\circ$. 
In a large-scale view (e.g., Howard et al. 1997) it is clearly seen that the 
two brightest YSOs dominate the center of a cluster much more extended 
than shown in the NICMOS image.

A substantial number of faint cluster stars are seen in Figure 2 
in addition to the brighter cluster stars measured by 
Tamura et al. (1991) and Itoh et al. (2001). 
The positions of all stars and any identifying names (e.g., 2MASS) are 
shown in Figure 2c and given in Table~3.
We note that one of the objects, Star 3 (NIRS11), is not a point source but 
instead is nebulous, like an embedded low-mass YSO.
The absolute coordinates were determined by assuming that NIRS3 
is located at the position of the compact radio source seen by 
Rengarajan \& Ho (1996), which position has an uncertainty $\sim 0.1''$.
Because of confusion due to both nebulosity and multiple stars, 
only a few of the stars measured by 2MASS 
have well determined coordinates, with nominal uncertainties $\lesssim 0.16''$. 
However, the range of differences between our positions and the 2MASS positions is $>0.16''$, 
with the discrepancies in both the positive and negative directions
(with the range justifying our assumption of the coincidence of NIRS3 
and the Rengarajan \& Ho source), 
and so we conclude that the absolute coordinates have uncertainties of 
several tenths of an arcsecond.
The relative coordinates, however, have uncertainties of only $\sim 0.02''$,
thanks to NICMOS's $0.076''$ pixel size and $\sim 0.2''$ diffraction limited 
resolution.
There are additional objects in the 2MASS Point Source Catalog that are within our field of view;
however, these are not stars but are some of the brighter clumps of nebulosity 
seen in Figure~2. 
We note that the {\it Spitzer} IRAC images of the nebulosity and the center of the cluster 
are saturated in all wavelength bands.

The polarization, position angles, and estimated magnitudes at 2.0 \micron\ are also
given in Table 3,
where the NIRS identifying numbers are from Tamura et al. (1991), Miralles et al. (1997),
and Itoh et al. (2001). 
The polarization vectors for those stars with good measurements are plotted in Figure 2c
(to be ``good'', it is required that the measured $P/\sigma_P > 5$ and also 
that the star can be detected on the $Q$ and $U$ images).
Omitting NIRS1, NIRS3, and the YSO because their polarization is strongly affected 
by scattering, the average $\theta$ for all these stars is $\theta \sim 134^\circ \pm 7^\circ$.
Even though at least some of the stars in this region of star formation 
are probably YSOs embedded in dense envelopes, 
which could produce polarization by scattering, 
the fact that most of the polarization vectors have more or less the same position angle 
(Table 3)
leads us to infer that the main cause of the polarization is 
dichroic absorption by elongated grains aligned by the interstellar magnetic field.
Since the grain rotation axis is parallel to the magnetic field direction, 
we infer that the direction of the field in the plane of the sky 
is $\sim 134^\circ$ east from north.

In Figure~2 we see that the scattered light lobes from NIRS1 
are much more extended than had previously been observed 
(Tamura et al. 1991; Miralles et al. 1997; Itoh et al. 2001), 
not only to the north but also to the south 
where an extension of the NIRS1 scattered light lobe 
lies south of the southwest lobe illuminated by NIRS3.
This is most obvious in polarized light, which we plot in Figure 5.
Tamura et al. (1991) first suggested that the northeast and southwest lobes, 
which they called IRN 1 (infrared reflection nebula 1), are illuminated by NIRS3.
They labeled all the polarized nebulosity surrounding NIRS1 as IRN 2.
With the significantly better resolution and sensitivity of NICMOS, 
we see that both NIRS3 and NIRS1 illuminate their own scattering light lobes, 
and that the south lobe of NIRS1 is visible at least $8'' - 9''$ south of NIRS1,
that is, well south of the bottom of the southwest lobe of NIRS3.
In Figure 5 we have drawn straight lines through NIRS1 and NIRS3 
to show the approximate centers of the two scattering lobes of each star, 
including the extension of the south lobe of NIRS1 below the southwest lobe of NIRS3,
located south of RA, Dec ($2.5'',-5''$).
The transition from the polarization position angle indicating illumination by NIRS3 
to the polarization position angle indicating illumination by NIRS1 
is very distinct at this location  
on a plot of the polarization position angle image. 
We conclude that the south scattered light lobe of NIRS1 lies 
behind NIRS3's southwest lobe on the plane of the sky 
and the lobes may not interact at all.
The north lobe of NIRS1 is probably aimed substantially towards the earth 
(its maximum polarization is only 44\%) 
but the two lobes of NIRS3 may lie close to the plane of the sky
(the maximum polarization of the northeast lobe is 80\%).
With a separation on the sky of $\geq 6000$~AU, it is most unlikely that NIRS1 and NIRS3 
form a binary pair.

The polarized light is scattered from dust in the vicinity of the star;
usually this dust is found in the outflow itself or in entrained gas and dust in the outflow.
We can use the location of the observed scattered light lobe to infer 
the geometry of the YSO's outflows. 
In Figure 5 we see that maxima of the scattered light intensities for NIRS1 
do not follow the straight lines that mark
the center of the scattered light lobes of NIRS1 but 
instead are bent into a big ``S'' curve, as marked by the dashed curves.
Although it is possible that these maxima of scattered light intensities are 
simply clumps of dust irregularly and randomly positioned in the outflow, 
it is also possible that these maxima mark
the direction of the outflows from the two YSOs. 
If this is the case, we can infer that 
the direction of the most recent outflow ejection from NIRS1 does not lie 
along the axis of the overall outflow.
This is the appearance of an outflow that is not exactly aligned with the rotation axis 
of the star and disk 
such that the outflow direction precesses around the rotation axis.
Alternatively, NIRS1 is a binary where the rotation axis of the star producing the outflow 
precesses around the orbital axis of the system.

A number of YSOs (e.g., IRAS 20126+4104, Shepherd et al. 2000) 
have now been observed to have outflows with bends or twists.
It is generally thought that the most likely explanation is that the YSO is part of 
a binary whose rotation axes are misaligned with the orbit  
(see, e.g., Monin et al. 2007 for a review).
Bate et al. (2000) and Terquem et al. (1999) discuss the relevant equations 
and give examples for systems where the presence of  
a companion star in a non-coplanar orbit induces precession in the 
axis of the observed star's disk. 
We suggest that a similar binary configuration is the cause of the twists in the scattered 
light lobes of NIRS1. 
A separation $< 500$ AU would be necessary for the binary, 
given the 500 AU resolution of NICMOS for the 2.4 kpc distance of S255 
and the fact that NIRS1 appears slightly larger than a point source to NICMOS.

\subsubsection{Source Parameters}

NIRS3 is a much more luminous YSO than NIRS1.
The observed fluxes for the two YSOs are plotted in Figure 6, 
where the J, H, K, L$'$, and M$'$ fluxes were measured by Itoh et al. (2001) on Subaru 
and the 7 -- 20 \micron\ fluxes were measured by Longmore et al. (2006) 
on Gemini North. 
Fits consisting of two graybodies of different temperatures were made for each YSO; 
the least-squares fitted parameters are given in Table~4.
The extinction law used in the fit was that of Draine (2003a, 2003b) for $R_V = 3.1$.
The stellar masses and effective temperatures 
were estimated by interpolating the measured luminosities 
in the ``canonical'' pre-main-sequence evolutionary models of Bernasconi \& Maeder (1996)
for both the luminosities at the zero-age main sequence (ZAMS) 
and the luminosities at the maximum luminosity on the radiative track, 
where the luminosity is due to both contraction and nuclear burning.
The ZAMS spectral types for stars with the luminosities of 
NIRS1 and NIRS3 are approximately B3 and B1, respectively. 
The numbers of ionizing photons, $Q_0$, were then estimated from the 
dwarf (log $g = 4$) model stellar atmospheres of Sternberg et al. (2003)
for ZAMS stars of the measured luminosities (Table~4).
Since there is no measured radio flux for NIRS1, 
we estimate that it has essentially no ionizing photons, which is reasonable for a B3 star.
For NIRS3 we estimate $Q_0 \sim 1.24 - 2.8 \times 10^{45}$ s$^{-1}$  
from the radio fluxes  $F_\nu = 1.97$ mJy and 4.4 mJy  
measured by Rengarajan \& Ho (1996) and Snell \& Bally (1986), respectively, 
and equation 1 of Simpson \& Rubin (1990).
These fluxes of ionizing photons are too small for a B1 star (Table~4) --- this is 
another indication that NIRS3 is still pre-main sequence. 

The temperatures of the two components in Table 4 are obviously not the effective 
temperatures, $T_{\rm eff}$, of the stars themselves but are the temperatures of 
the warm and cooler dust in the optically thick, dusty envelopes or disks 
obscuring the stars.
The orientation of the disk and outflow also affect the SED as viewed from the Earth.
The Online SED Model Fitter of Robitaille et al. (2007) 
can also be used to estimate a YSO's luminosity, inclination, stellar $T_{\rm eff}$, 
and evolutionary state from its observed MIR and far-infrared SED.
Using the Online SED Model Fitter we find that results for the extinction, luminosity, 
and YSO mass are similar to those of Table~4, although $T_{\rm eff}$ is not 
well constrained, probably because the wavelengths of the NIR observations 
are all on the Rayleigh-Jeans part of the stellar SED
and there was no input for the disk inclination.
Because of this good agreement, we conclude that the models of Table~4 
are good estimates of the extinction and luminosities of  
the dusty envelopes, given the available data, 
and enable us to estimate the masses and spectral types of NIRS1 and NIRS3.

\subsubsection{Parallel and Non-parallel Polarization Patterns}

Jiang et al. (2008) measured the K-band (2.2 \micron) polarization around NIRS1 with 
the adaptive optics system on Subaru (resolution $0.15'' - 0.2''$).
They observe that there is a lane of low polarization with position angle $\sim 110^\circ$
in the vicinity of NIRS1  
and that the polarization vectors in the lane are  
parallel to the lane and perpendicular to the scattered light lobes. 
They suggest that the cause of this ``polarization disk'' 
is that there is an edge-on, optically thick disk or toroid surrounding NIRS1.
However, as we demonstrate below, this parallel polarization pattern is 
due to instrumental scattering and telescope diffraction of 
the polarized light of NIRS1 ($P = 24$\%, Table 3).

A telescope, through diffraction and instrumental scattering, 
can introduce substantial diffuse polarized light 
to an image, which can extend to many arcseconds distance from a source, 
if the source is itself polarized (e.g., Simpson et al. 2006).
This production of a parallel polarization pattern by beam smearing was first noted 
by Whitney et al. (1997) and was also described by Lucas \& Roche (1998).
To minimize the production of an artificial polarization disk, 
we produced the polarization images a second time, subtracting the PSFs for NIRS1 and NIRS3, 
which we estimated by scaling the individual POL images of our red standard star, Oph-N9. 
The PSFs were subtracted before calculating the polarization 
and rotating to celestial coordinates. 
This PSF subtraction results in significant differences 
in the appearance and observed polarization of the remaining diffuse emission. 
After PSF subtraction, the parallel polarization vectors of the parallel polarization pattern 
are no longer obvious. 
To show this, we plot the area around NIRS1 and NIRS3 in more detail in Figure~7.
The parallel polarization pattern on either side of NIRS1 is no longer apparent 
and the region of low polarization is on one side of the star only. 

We conclude that for the case of NIRS1, the apparent parallel polarization pattern 
is due to diffraction and instrumental scattering 
and thus there is no evidence for a ``polarization disk'' in this source.
We pointed out earlier (e.g., Fig. 3b) that the presence  
of regions of low polarization or parallel polarization vectors 
is not evidence for a physical disk although it is evidence for 
an optically thick toroidal structure in the YSO's envelope.
A small disk with diameter less than the telescope resolution could be present --- 
the YSO would look like a point source but could have substantial polarization 
from scattering by the disk. 

There are other asymmetries in the appearance of NIRS1 in Figure~7.
There is still a region of lower polarization $\sim 0.5''$ south of NIRS1 
and there are also highly polarized, brightish regions $0.25'' - 0.35''$ northwest of NIRS1.
These polarized features within $0.35''$ ($\sim 800$~AU) of NIRS1 are reminiscent of 
the flared disks in the models of Whitney \& Hartmann (1992) and Stark et al. (2006).
The axis of the disk would be tipped toward the Earth. 
On the other hand, clumps of dust in the outflow cavity could produce an asymmetric appearance, 
as could aligned grains in a twisted magnetic field geometry (Whitney et al. 2009, in preparation).
The small NW feature does not look like a star, since it is much fainter 
and it is visible in only two of the three POL filters in both measurements of the source; 
however, the feature also coincides with structure in the first Airy bright ring  
of the NICMOS diffraction pattern (cf., NIRS3 in Fig.~7a), 
and thus we cannot exclude the possibility that it is an extremely polarized star.

We next consider the relation between the magnetic field direction, 
the outflow directions, and the polarization position angles of the stars 
producing the outflows. 
We assume that the interstellar dust grains are aligned by the magnetic field 
and that we can therefore infer its direction from the polarization position angles 
of the field stars (see the review of dust grain alignment by Lazarian 2007). 
Originally it was suggested for low-mass star formation that the cloud of gas 
and dust collapses along the magnetic field lines to form a disk around the new star
(e.g., Shu et al. 1987).
For such a case, the outflows would be perpendicular to the disk 
or approximately parallel to the magnetic field. 
However, it is commonly observed that YSOs, especially those that appear as point sources, 
have polarization position angles perpendicular to the outflows as defined by 
their scattered light lobes or CO outflows --- that is, 
their polarization position angles are parallel to their disks.
Models have shown that dense envelopes or disks and bipolar cavities 
can produce such polarization patterns through multiple scattering 
(e.g., Fig. 3b; Whitney \& Hartmann 1993; Lucas \& Roche 1998; Bastien \& M\'{e}nard 1988).
This is probably the case for NIRS1, where its polarization position angle 
is approximately perpendicular to its scattering lobes but is not well-aligned 
with the local magnetic field as determined from the other cluster stars (Figs. 2 and 5). 

The BN object in Orion is another case where its polarization position angle is 
almost identical to that of the local magnetic field as estimated from the other stars in the 
field of view and also approximately perpendicular to its outflow direction 
(Simpson et al. 2006; Jiang et al. 2005).
Probably both dichroic absorption and scattering contribute 
to its extremely high 29\% polarization at 2 \micron\ in BN's central core. 
BN is also similar to NIRS1 in that PSF subtraction is necessary 
to see the details of any polarization structure near the star 
(Simpson et al. 2006).
Although one sees a line of low polarization apparently extending through BN, 
this line of low polarization is not due to a 
polarization disk as was claimed by Jiang et al. (2005). 
Instead, it is due to a dust lane stretching over 10,000 AU across the region 
(Simpson et al. 2006).

Massive stars, however, do not necessarily have envelopes that collapse along 
magnetic field lines (e.g., Zinnecker \& Yorke 2007 and references therein).
We can test this by considering the relative angles of polarization 
and outflow direction. 
Whitney \& Wolff (2002) and Lucas (2003)
showed that if the optical depth in the envelope is high enough
and the grains are elongated, 
the polarization is dominated by dichroic absorption.
For this case, the polarization can be large 
and the polarization position angle is parallel to the magnetic field 
if the grains in a YSO envelope are aligned by the magnetic field.
A number of massive YSOs have been found where the magnetic field 
is not aligned with the observed molecular outflow direction,
for example by Aitken et al. (1993), as inferred from MIR polarimetry.
With beam sizes of $4''$ to $6''$, these measurements included almost the whole envelope.
We next discuss a NIR example, NIRS3, which is a point source to NICMOS 
(source diameter $< 500$ AU). 

NIRS3 has scattered light lobes with position angles $\sim 67^\circ$ and $\sim 231^\circ$
(Fig. 5) and a polarization position angle of $\sim 27^\circ$ (Table 3).
The polarization position angle observed for NIRS3 is aligned with neither the magnetic field 
nor with the direction of its scattered light lobes (Table 3 and Fig. 2).
This is surprising since the $\sim 134^\circ$ position angle of the inferred magnetic field 
is almost perpendicular to the position angles of the scattered light lobes and therefore 
approximately parallel to any disk NIRS3 might have. 
We would expect to see a situation like that of BN 
where the polarization position angle is approximately the same as the inferred magnetic field 
and perpendicular to the outflow direction (Simpson et al. 2006; Jiang et al. 2005). 
Like BN, NIRS3 has a substantial optical depth, $\tau_{2.0 \mu m}$, at 2.0 \micron\ 
($\tau_{2.0 \mu m} \sim 5.5$, see Table 4); 
this is seen by the depth of the silicate feature at 10 \micron\ and 
by the steep SED in the NIR (Fig. 6, Table 4).
This $\tau_{2.0 \mu m}$ could likewise be due to a modest sized disk (radius $\lesssim 240$ AU) 
without the source appearing extended since 
the NICMOS resolution of $0.2''$ corresponds to 480 AU at the S255 distance of 2.4 kpc. 

We consider whether the observed polarization position angle in NIRS3 ($\theta = 27^\circ$) 
is, in fact, some other intrinsic angle rotated 
by dichroic absorption from dust grains either along the line of sight 
or in the NIRS3 envelope 
which are aligned by the intervening magnetic field.
It is not possible to rotate a polarization position angle 
corresponding to the disk (which would be perpendicular to the scattered light lobes 
with position angles $\sim 51^\circ$ and $\sim 67^\circ$)
to the observed position angle seen in NIRS3.
However, if the magnetic field position angle is $\sim 150^\circ$ (rather than 
the $\sim 134^\circ$ estimated from the stars), 
a polarization position angle {\it parallel} to the scattered light lobes 
can be rotated to the observed position angle.
The amount of foreground polarization subtracted would need to be $\sim 15 - 20$\%
(the equations are given in the Appendix).
The resulting intrinsic polarization of NIRS3 is then $\gtrsim 22$\%.
This amount of rotation is not inconceivable given that $\tau_{2.0 \mu m} \sim 5.5$.
Since the intrinsic polarization position angle would be parallel to the 
scattered light lobes instead of the more usual position angle 
perpendicular to the lobes (due to scattering), 
it could be a case of a massive YSO with a magnetic field substantially different 
from that of the local magnetic field. 
A test for polarization rotation (e.g., Martin 1974; Appendix)  is the following: 
if there is rotation of an intrinsic polarization angle due 
to dichroic absorption, NIRS3 should also exhibit  
observable circular polarization.
A number of objects have now been observed to exhibit both linear and circular NIR polarization 
(e.g., Gledhill et al. 1996; Chrysostomou et al. 2000; Clark et al. 2000).

The above suggestion of rotation of the polarization position angle by foreground 
dichroic absorption may, in fact, not be necessary for NIRS3.
We recall that Aitken et al. (1993) has found a number of massive YSOs whose 
envelope magnetic fields, as determined by MIR polarimetry, 
are misaligned with the disk and/or outflow directions.
The low-mass YSO, HL Tau, also appears to have a twisted magnetic field misaligned 
with its outflow axis (Lucas et al. 2004). 
This was determined by NIR polarization measurements combined with 
a detailed Monte-Carlo scattering model. 
Moreover, Whitney et al. (2009, in preparation) have found that models 
of YSOs with twisted magnetic fields and aligned grains can have the overall polarization 
misaligned with both the disk/envelope plane and the outflow axis.
These models also have substantial circular polarization.
NIRS3 may also be a member of this class of YSO.

Finally, we discuss the fact that 
the line connecting NIRS3's two scattered light lobes is not straight (Fig. 5).
There are two possibilities that might be occurring: 
(1) NIRS3 is moving to the north or northwest, and only from time to time does it 
eject enough material along its axes such that cavities producing bipolar scattered light lobes 
become visible, or 
(2) The scattered light lobes actually are cones which have the ``X'' configuration seen in models 
such as those of Stark et al. (2006), where the edges of the cones produce more 
scattered light because they have longer path lengths, 
but the north side of the cones is either obscured by additional dust 
or it has been disrupted by interaction with other stars in the cluster.
However, there is no indication of this needed additional extinction in the 
larger scale images of S255 of Howard et al. (1997) or Miralles et al. (1997).
Consequently, we prefer the idea that the apparent bend in the scattered light lobes 
is due to the episodic ejection of matter by NIRS3.

\section{Summary and Conclusions}

We have imaged the YSOs NGC 6334 V and S255 IRS1 
with the NICMOS Camera 2 polarization filters on {\it HST}
at $0.2''$ resolution ($< 350$ AU or $< 500$ AU, respectively ).
NGC 6334 V consists of a deeply embedded massive YSO 
with bipolar outflows in the plane of the sky.
Only the outflow regions can be detected at 2 \micron, 
where they are seen as regions of highly polarized, scattered light.
The star itself is detected at MIR wavelengths, such as 
the 5.6 and 8.0 \micron\ Spitzer IRAC images.
There is a very compact radio source at the location of the illuminating star 
(Rengarajan \& Ho 1996);
two other radio sources are located in the east outflow region.
We speculate that these are due to photons from the massive YSO 
ionizing gas along the outflow axis.

The illuminating star of the bipolar lobes of NGC 6334 V lies 
significantly ($2''$) south of the line connecting the two lobes.
On the other hand, the peak of the 20 \micron\ flux lies further north, 
along the line connecting the lobes (Simon et al. 1985; Kraemer et al. 1999).
We suggest that there is a second, less evolved and more luminous YSO at this location, 
which is too deeply embedded to be detected at any wavelength shorter than $\sim 8$ \micron.
This second YSO produces the bipolar molecular outflow oriented approximately in the line of sight 
whereas the first YSO produces the NIR bipolar lobes as well as illuminates them.
The asymmetry in the position of the illuminating star with respect to the 
scattered light lobes is due to extinction.

S255 IRS1 contains two YSOs, separated by $\sim 2.5''$ and 
known as NIRS1 and NIRS3 from the imaging of Tamura et al. (1991).
Both YSOs are visible at 2 \micron, but both are heavily extincted, especially NIRS3,
the more luminous star.
Each YSO illuminates two lobes of polarized scattered light; 
the lobes of one star are at an angle to those of the other, such that 
the southern lobe  of NIRS1 is partially obscured by the southwest lobe of NIRS3.

Unlike the region around NGC 6334 V, which is obscured by a dense dust cloud, 
there is a large number of other stars in a cluster surrounding NIRS1 and NIRS3. From 
the measured polarization of these stars we infer 
a magnetic field direction of $\sim 134^\circ$ in the S255 IRS1 region. 
The polarization position angle of the core of NIRS1 is more or less aligned with this field. 
It is, in fact, reasonably perpendicular to the direction of the parts of the scattered light 
lobes that are closest to the YSO.
The location of the center of the NIRS1 scattered light lobes changes symmetrically 
with distance from the YSO --- 
we conclude that the outflow from S255 NIRS1 appears to precess around an axis.
Such precession probably indicates that NIRS1 is part of a binary with separation 
$\lesssim 800$ AU. 

The polarization direction of NIRS3 is $\sim 27^\circ$, which corresponds to 
neither the direction of the scattered light lobes nor to their perpendiculars.
We show that it is possible that the polarization vector that would be seen in 
the vicinity of NIRS3 has been rotated to the observed position by 
dichroic absorption from dust grains aligned by the magnetic field between us and the YSO. 
This can be tested by measuring the circular polarization of NIRS3.
On the other hand, NIRS3 may be another example of a massive YSO 
where the inferred magnetic field direction is aligned with neither the outflow axis 
nor the disk plane (e.g., Aitken et al. 1993).

The illuminating star of the bipolar lobes of S255 NIRS3 lies 
significantly ($\sim 1''$, or $\sim 2400$ AU) above the line connecting the two lobes.
We suggest that NIRS3 is moving to the northwest, thereby causing 
the misalignment between the axes of the lobes.  

We conclude that none of these three massive YSOs has a simple morphology 
of an accretion disk and outflow. 
The scattered light lobes of both NGC 6334 V and S255 NIRS3 are separated by several arcseconds  
from the YSOs themselves; from this we conclude that the ejection is episodic.
Moreover, both of these YSOs are located $\sim 1'' - 2''$ from the line joining their 
scattered light lobes, possibly indicating movement between epochs of mass ejection. 
Finally, S255 NIRS1 appears to have an asymmetric disk that precesses.
Most likely, at least two and possibly all three of these massive YSOs are close binaries,
with the misalignment of the binaries responsible for the observed asymmetries
in their outflow structures.

\acknowledgments
Support for program 10519 was provided by NASA through a grant from 
the Space Telescope Science Institute, which is operated by the Association of Universities 
for Research in Astronomy, under NASA contract NAS5-26555.
JPS acknowledges support from NASA/Ames Research Center Research Interchange Grant NNA05CS33A
to the SETI Institute.
This publication makes use of data products from the Two Micron All Sky Survey, 
which is a joint project of the University of Massachusetts and the Infrared Processing 
and Analysis Center/California Institute of Technology, funded by the 
National Aeronautics and Space Administration and the National Science Foundation.
This work is based in part on observations made with
the Spitzer Space Telescope, which is operated by the Jet Propulsion
Laboratory, California Institute of Technology under NASA contract 1407.
We thank K. Kraemer for helpful discussions. 
We thank the referee for the thoughtful comments and suggestions 
that improved the presentation. 

{\it Facilities:} \facility{HST (NICMOS)}, \facility{Spitzer (IRAC)}.

\appendix
\section{Dichroic Screen Subtraction}

Radiative transfer of polarized light has been discussed by a number of authors, 
most recently by Martin (1974), Jones (1989), and Whitney \& Wolff (2002).
Here we compute the polarization due to a slab of dust lying between a polarized YSO 
and the Earth, where the grains in the slab are aligned at a different angle 
from  that of the source polarization position angle.
Our goal is to determine the source polarization ($I_0, Q_0, U_0, V_0$) by subtracting the change 
caused by the dust slab from the observed ($I, Q, U$) and possibly $V$. 
We start by writing the radiative transfer equations A13 and A14 of Martin (1974),
which assume linear dichroism and birefringence only,  
in the notation of Jones (1979):
\begin{mathletters}
\begin{eqnarray}
{dI \over ds} & = & -\kappa I +  \kappa_P Q \cos 2\theta + \kappa_P U \sin 2\theta \\
{dQ \over ds} & = & \kappa_P I \cos 2\theta -  \kappa Q  \ \ \ \ \  + \Delta\epsilon V \sin 2\theta \\
{dU \over ds} & = & \kappa_P I \sin 2\theta \ \ \ \ \ - \kappa U  - \Delta\epsilon V \cos 2\theta \\
{dV \over ds} & = & \ \ \ \ - \Delta\epsilon Q \sin 2\theta + \Delta\epsilon U \cos 2\theta - \kappa V
\end{eqnarray}
\end{mathletters}
where $\theta$ is the polarization position angle (east from north) of the slab dust,
$\kappa$ is the average of the absorption coefficients parallel and perpendicular 
to the direction of grain alignment, $\kappa_P$ is the difference of the parallel 
and perpendicular absorption coefficients divided by 2
(in the notation of Martin 1974, $\Delta\sigma = \sigma_1 - \sigma_2 = \kappa_P$), and 
$\Delta\epsilon$ is the coefficient for linear birefringence, to be discussed later.

We assume that the polarized source has components $I_0$, $Q_0$, and $U_0$, but 
$V_0 = 0$. 
Whitney \& Wolff (2002) showed that YSOs with aligned grains in their dusty envelopes 
can produce copious circularly polarized light; however, if the source is axisymmetric 
and unresolved, the positive and negative circular polarizations cancel and the 
net circular polarization is zero. 
(We note that if the magnetic field geometry  is not axisymmetric, 
for example if it has a toroidal component,
then the net circular polarization is not zero, Whitney et al. 2009, in preparation.) 
With $V_0 = 0$, the first three equations are the same as those of Jones (1989), 
and we use his solution for the $I$, $Q$, and $U$ Stokes parameters  
for a slab of constant dust properties:
\begin{mathletters}
\begin{eqnarray}
I & = & (I_0 \cosh \tau_P + \zeta \sinh \tau_P) e^{-\tau} \\
Q & = & (I_P \cos 2\theta + Q_0) e^{-\tau} \\
U & = & (I_P \sin 2\theta + U_0) e^{-\tau}
\end{eqnarray}
\end{mathletters}
where $\tau = \int \kappa ds$, $\tau_P = \int \kappa_P ds$, 
$I_P = I_0 \sinh \tau_P + \zeta \cosh \tau_P - \zeta$, 
and $\zeta = (Q_0 \cos 2\theta + U_0 \sin 2\theta)$.

We suppose next that the dust slab  contributes a 
fractional polarization $P_s$ and $\theta$.
In the solution above, if $Q_0$ and $U_0$ are zero, then $P =P_s = \tanh \tau_P$. 
Writing the equations in matrix form we get 
\begin{equation}
\left(
\begin{array}{c}
I \\
Q \\
U 
\end{array}\right)
= e^{-\tau} 
\left(
\begin{array}{ccc}
\cosh \tau_P & \cos 2\theta \sinh \tau_P & \sin 2\theta \sinh \tau_P  \\
\cos 2\theta \sinh \tau_P & (\cos^2 2\theta (\cosh \tau_P-1)+1) & \sin 2\theta \cos 2\theta (\cosh \tau_P-1) \\
\sin 2\theta \sinh \tau_P & \sin 2\theta \cos 2\theta (\cosh \tau_P-1) &  (\sin^2 2\theta (\cosh \tau_P-1)+1) \\
\end{array}\right)
\left(
\begin{array}{c}
I_0 \\
Q_0 \\
U_0 
\end{array}\right)
\end{equation}
or
\begin{equation}
\left(
\begin{array}{c}
I \\
Q \\
U 
\end{array}\right)
= e^{-\tau} A 
\left(
\begin{array}{c}
I_0 \\
Q_0 \\
U_0 
\end{array}\right)
\end{equation}
where $A$ is the above matrix.
To solve this system for $I_0$, $Q_0$, and $U_0$, 
we estimate $\tau_P$ by iterating $\tau_P = P_s + \tau_P^3/3$ 
and multiply equation A4 by $e^{\tau} A^{-1}$.

We note that if we had expanded the sinh, cosh, and tanh expressions and used the first terms only, 
we would get equations 1--3 of Martin (1974) 
and the simplified equations used by Simpson et al. (2006).

The solution for the equation for circular polarization $V$ is derived by integrating 
equation 4 of Martin (1974):

\begin{equation}
{V \over I} = \tau_{\Delta\epsilon} ({U_0 \over I_0} \cos 2\theta - {Q_0 \over I_0} \sin 2\theta)
\end{equation}
where $\tau_{\Delta\epsilon} = \int \Delta\epsilon ds$.
Since to first order $\tau_P = P_s$, we can write 
$\tau_{\Delta\epsilon} = P_s \Delta\epsilon/\kappa_P$.
Martin (1974) gives expressions for the computation of $\Delta\sigma$ and 
$\Delta\epsilon$ and plots in his Figure 2 values of these quantities 
as a function of wavelength for two sets of refractive indices.
Although these almost certainly are not the refractive indices for 
real interstellar dust grains (e.g., Draine 2003a, 2003b),
we see that for the dielectric material with $m = 1.5 - 0.1i$,
the ratio $\Delta\epsilon/\Delta\sigma \sim 1$ at 2 \micron.
If this is the case for the dichroic screen subtraction that we describe in \S 3, 
we would predict that the screen produces a circular polarization $\sim 1$\%.

\clearpage

\begin{figure}
\epsscale{0.55}
\plotone{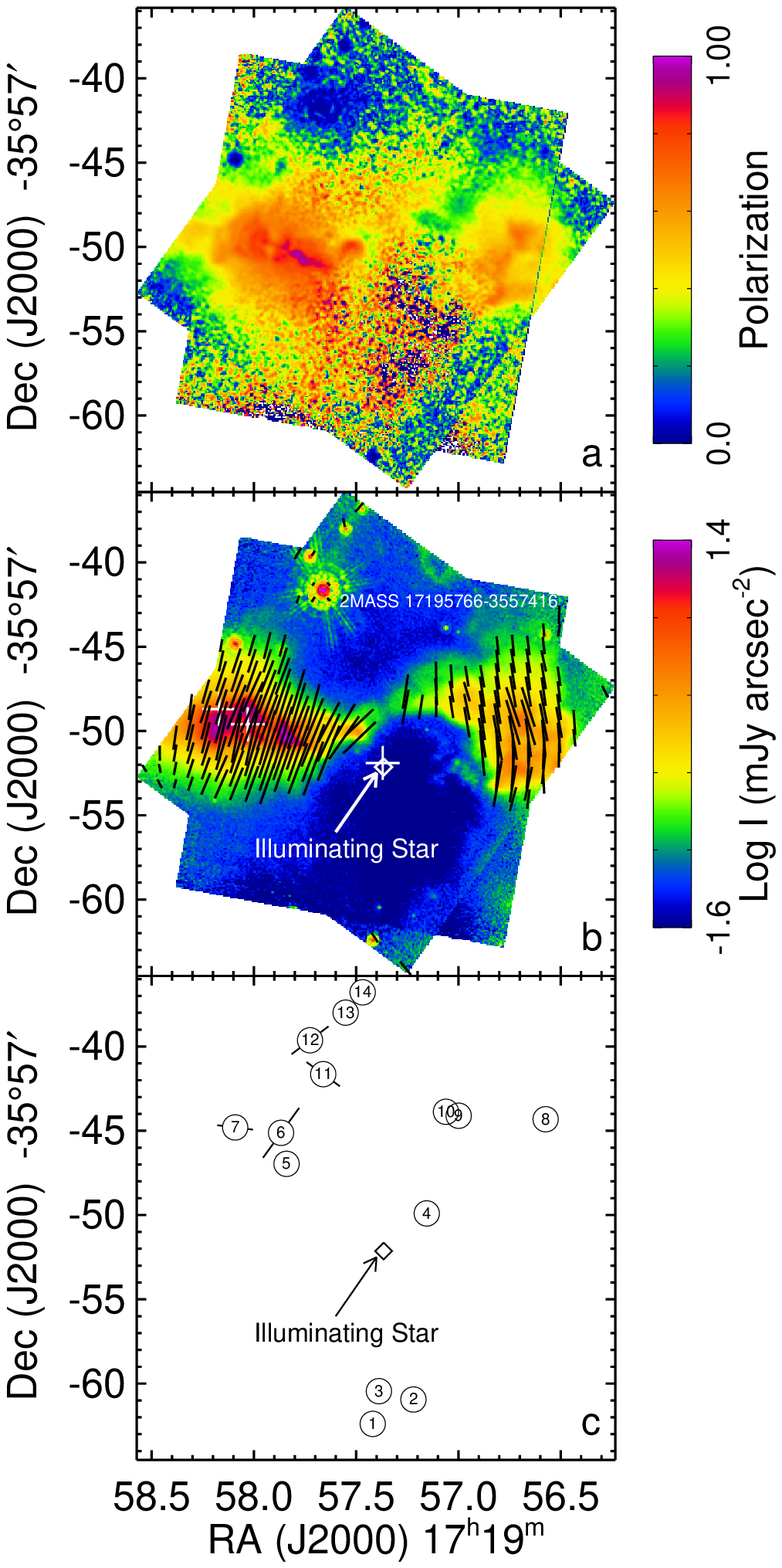}
\caption{
{\it HST} NICMOS image of NGC 6334 V. 
The absolute coordinates are derived from the location of the star 
2MASS17195766$-$3557416.
(a) Fractional polarization.
(b) Log intensity with polarization vectors.
The white crosses are the three radio sources (from left to right, R-E1, R-E2, and R-E3) 
detected by Rengarajan \& Ho (1996).
The white diamond is the location of the illuminating star as determined by a 
least squares fitting of the perpendiculars to the polarization vectors and
for all pixels with statistically significant polarized flux, $IP$.
(c) Locations of the stars detected with NICMOS (Table 2). 
The lines drawn through the stars with statistically significant polarization 
are proportional to $P$ plus a constant and plotted at the measured 
polarization position angle $\theta$. 
The diamond marks the location of the illuminating star. 
[{\it See the electronic edition of the Journal for a color version of this figure.}]
}
\end{figure}

\begin{figure}
\epsscale{0.55}
\plotone{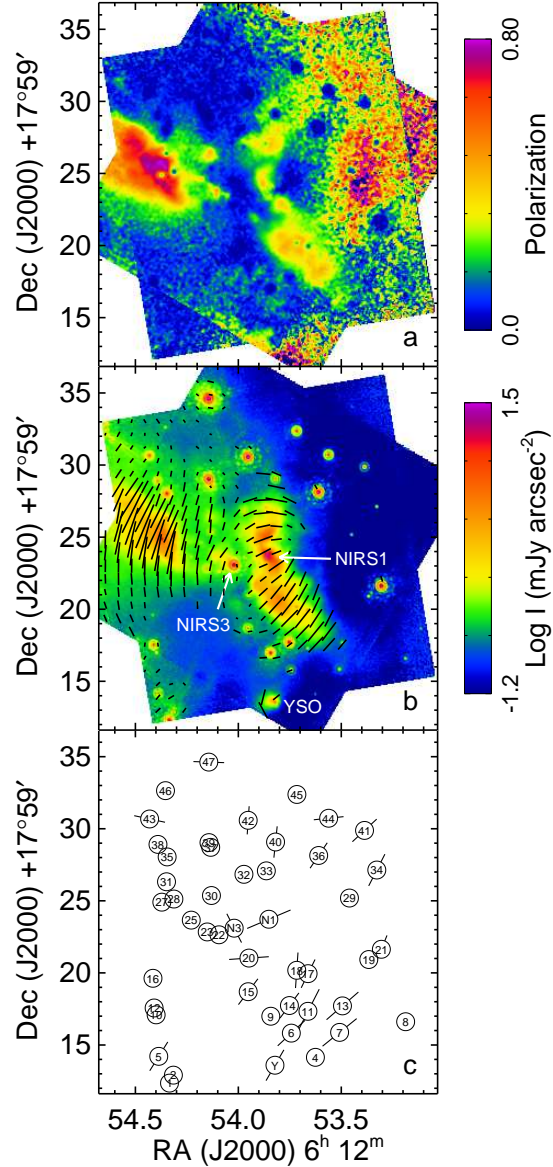}
\caption{
{\it HST} NICMOS image of S255 IRS1.
The coordinates are centered on the star NIRS3, 
which we assume to have RA = $6^{\rm h} 12^{\rm m}$54\fs02 and Dec = $+17$\degr 59\arcmin 23\farcs1 
(J2000) (Rengarajan \& Ho 1996; see text).
(a) Fractional polarization.
(b) Log Intensity with polarization vectors.
The locations of the stars NIRS1 and NIRS3 are marked, as is the location of a small 
extended source that is a newly discovered YSO.
(c) Locations of the stars detected with NICMOS (Table 3). 
The lines drawn through the stars with statistically significant polarization 
are proportional to $P$ plus a constant and plotted at the measured 
polarization position angle $\theta$. 
Sources NIRS1, NIRS3, and the YSO have been abbreviated ``N1'', ``N3'', and ``Y'', respectively.
[{\it See the electronic edition of the Journal for a color version of this figure.}]
}
\end{figure}

\begin{figure}
\epsscale{0.75}
\plotone{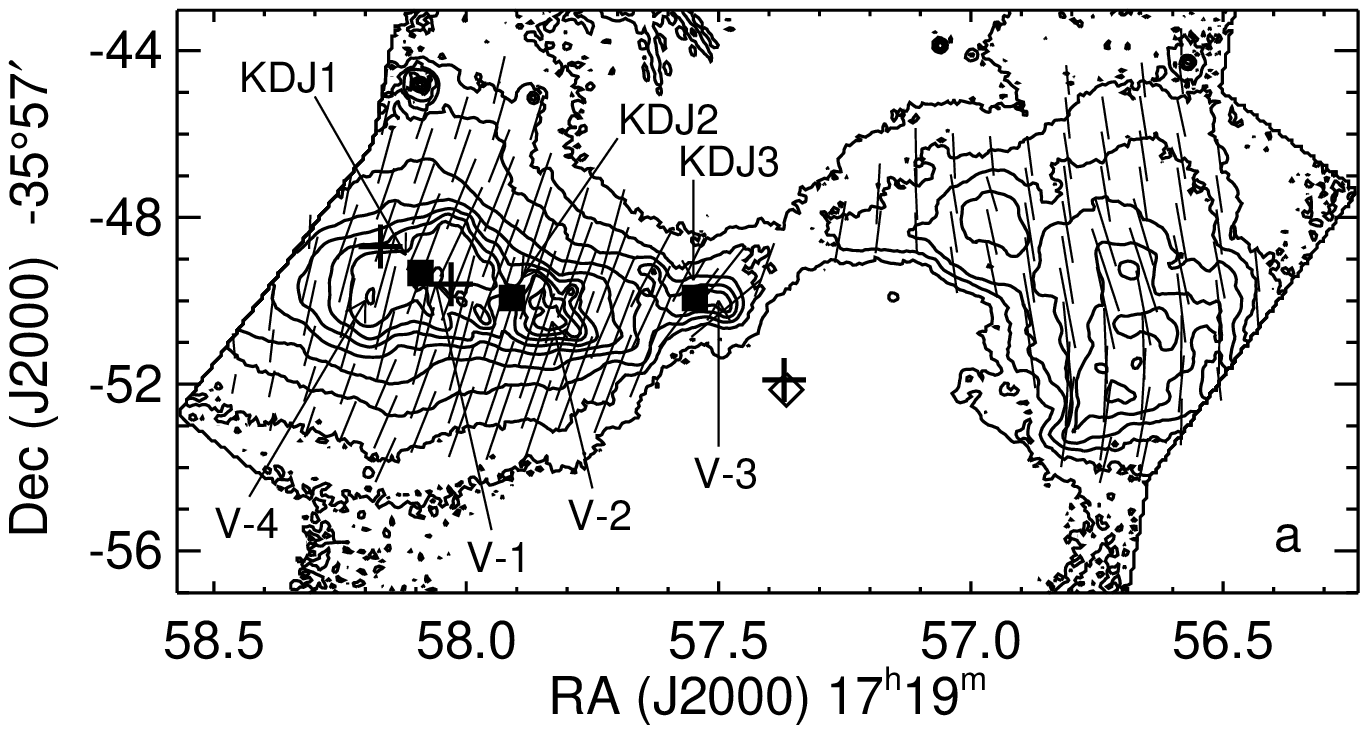}
\epsscale{0.64}
\plotone{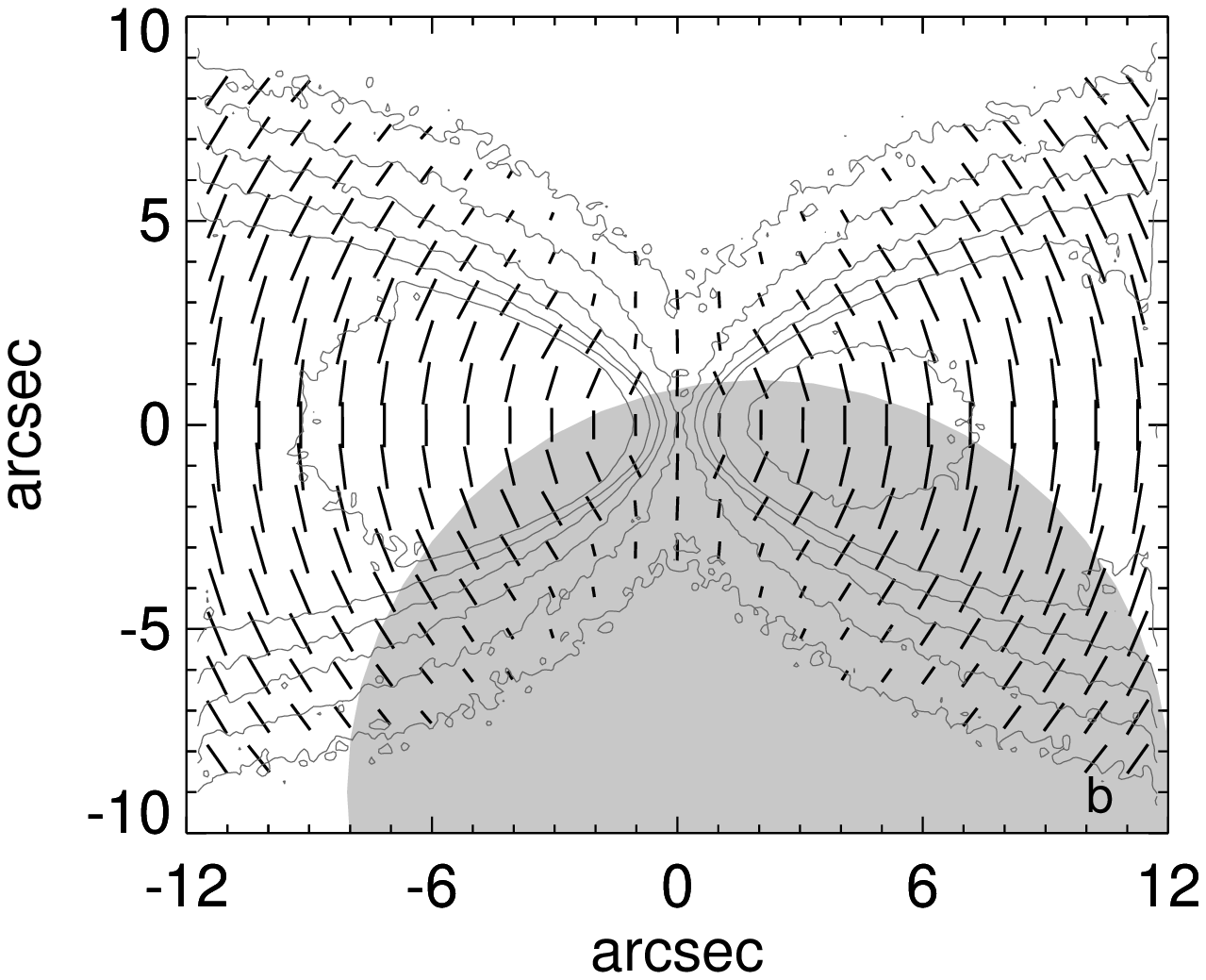}
\caption{
(a) Contours of the Stokes $I$ image of NGC 6334 V with polarization vectors. 
The crosses are the three radio sources detected by Rengarajan \& Ho (1996).
The diamond is the location of the illuminating star as discussed in the text.
The filled squares mark the locations of the MIR sources 
KDJ1, KDJ2, and KDJ3 
measured by Kraemer et al. (1999), positioned by assuming that 
their source KDJ4 is the illuminating star. 
The Stokes $I$ contours start at 0.0773 mJy arcsec$^{-2}$ and increase by a factor of 2 
for each succeeding contour to a maximum of 39.57 mJy arcsec$^{-2}$.
(b) Model of a massive YSO (see text) showing a parallel polarization pattern 
(``polarization disk''). 
The output from the Monte Carlo scattering program was convolved with a 
$0.2''$ (FWHM) Gaussian, which smooths the contours and slightly increases 
the size of the polarization disk.
Perpendiculars to the polarization vectors on the edges of the scattered light lobes 
point to the opposing lobe and not to the central star. 
We show how the center and the location of the central star could be blocked by 
an optically thick cloud (gray).
}
\end{figure}

\begin{figure}
\epsscale{0.8}
\plotone{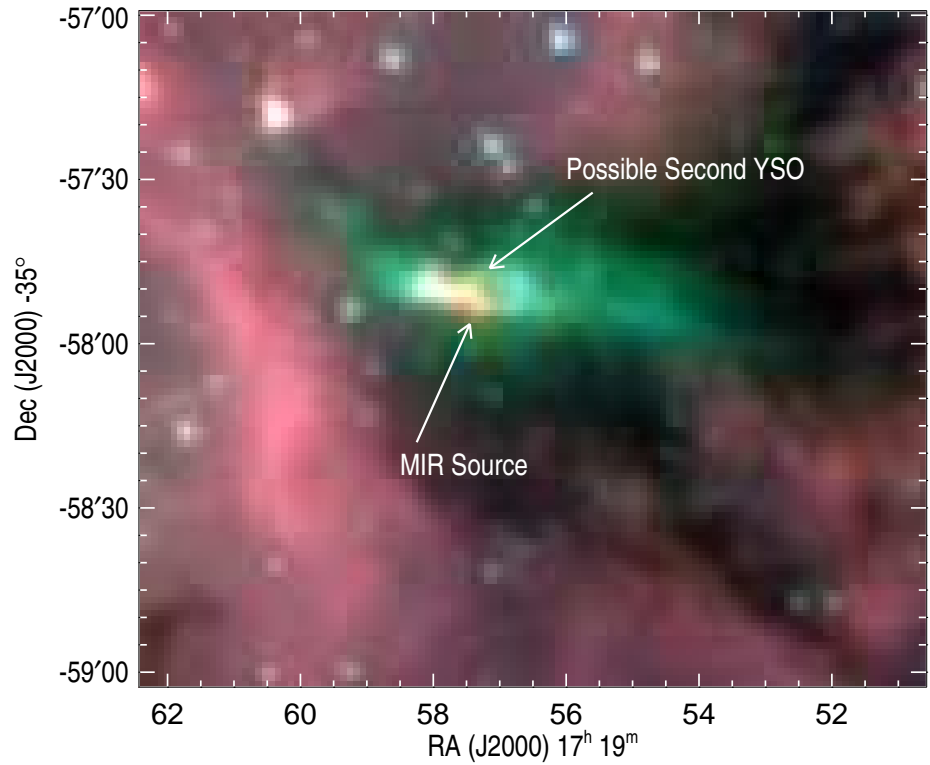}
\caption{
IRAC 3-color image of NGC 6334 V from the Spitzer GLIMPSE Legacy survey.
The blue, green, and red images 
are bands 1 (3.6 \micron), 2 (4.5 \micron), and 4 (8.0 \micron), respectively.
The location of the MIR source is marked, as is the location of a possible second YSO. 
}
\end{figure}

\begin{figure}
\epsscale{1.0}
\plotone{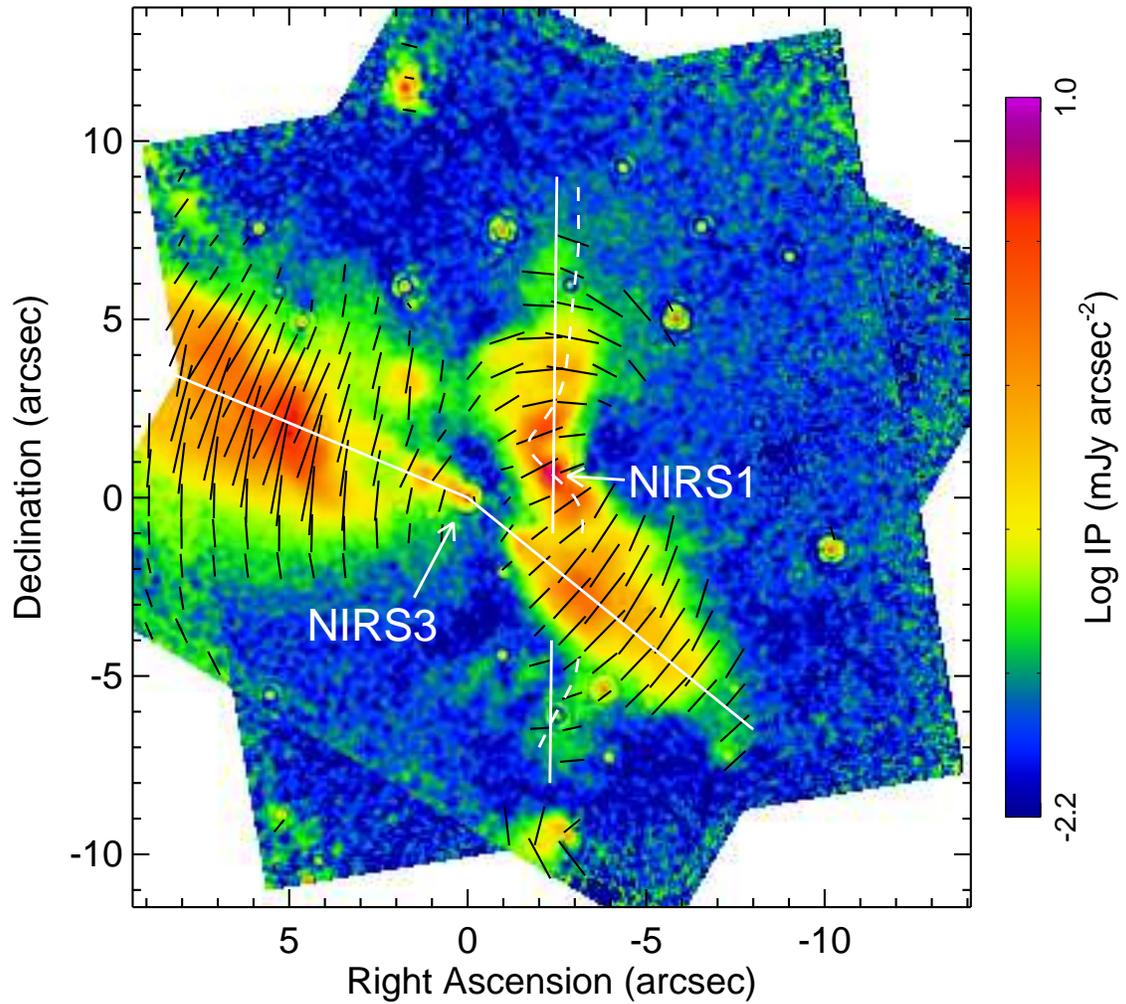}
\caption{
The polarized intensity image of S255 IRS1 with solid white lines 
marking the major axes of the lobes of scattered light.
The dashed white line follows (by eye) the location of the peak 
of the emission illuminated by NIRS1 in the north-south direction.
[{\it See the electronic edition of the Journal for a color version of this figure.}]
}
\end{figure}

\begin{figure}
\epsscale{1.0}
\plotone{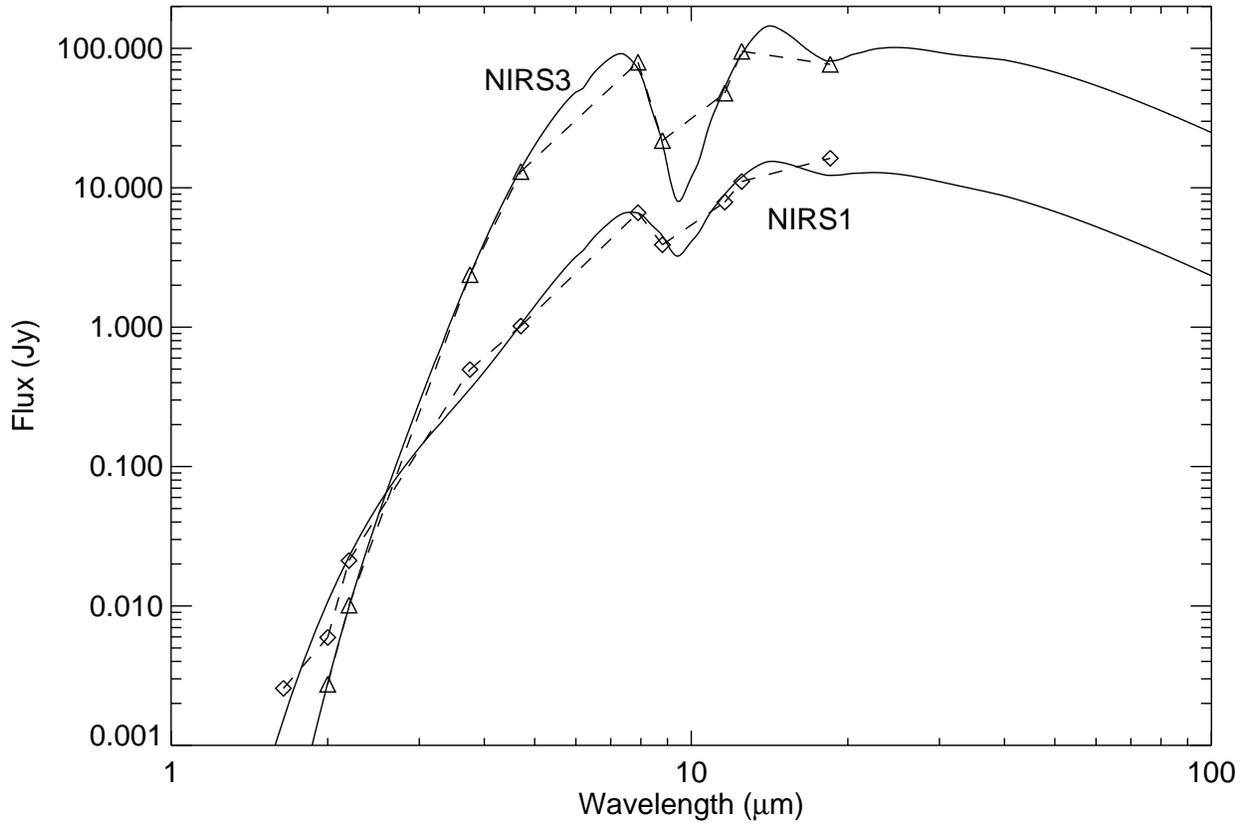}
\caption{
Observed fluxes for NIRS1 (diamonds) and NIRS3 (triangles) with fitted SEDs (see text).
The observed fluxes from 1 -- 6 \micron\ are from Itoh et al. (2001) 
with the exception of the 2.0 \micron\ points, which are from this paper, 
and the observed fluxes from 7 -- 20 \micron\ are from Longmore et al. (2006).
}
\end{figure}

\begin{figure}
\epsscale{1.0}
\plotone{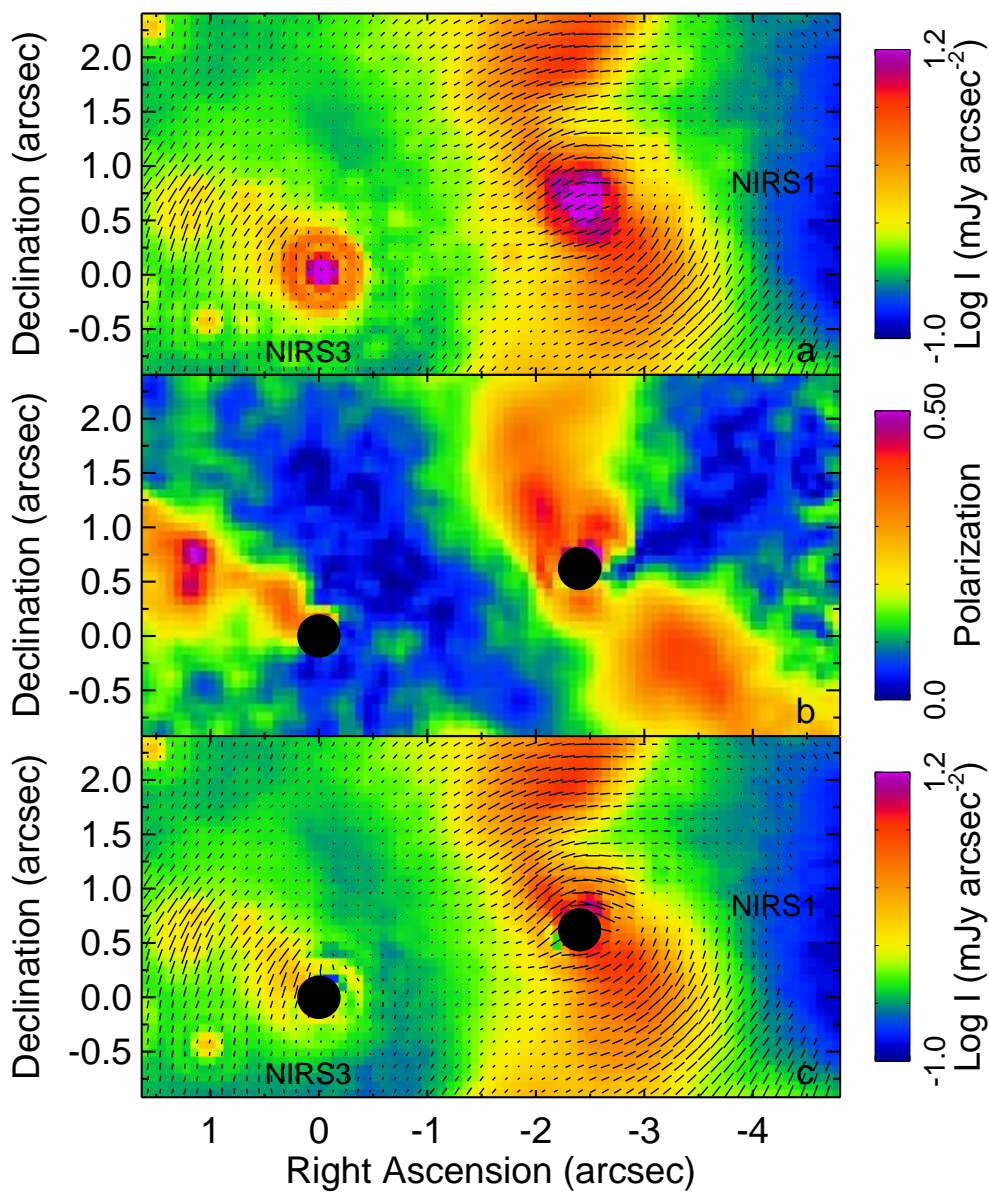}
\caption{
(a) An enlargement of the Stokes $I$ image from Fig. 2 showing the details of the area 
around NIRS1 and NIRS3.
(b) The polarization $P$ image after PSF subtraction.
To remove the effects of the PSF, scaled images of the standard star, Oph-N9, 
were subtracted from both NIRS1 and NIRS3 (see text).
The locations of NIRS1 and NIRS3 are indicated by large black dots.
(c) The Stokes $I$ image after PSF subtraction. 
[{\it See the electronic edition of the Journal for a color version of this figure.}]
}
\end{figure}

\clearpage

\input tab1.tex

\input tab2.tex

\input tab3.tex

\input tab4.tex

\end{document}

%% file: tab1.tex


\begin{deluxetable}{cccccc} 
\tablewidth{0pt} 
\tablecaption{Journal of the Observations}
\tablehead{ 
\colhead{Visit} & \colhead{Date} & \colhead{Target} & \colhead{Center RA} & \colhead{Center Dec} & \colhead{Camera 2} \\ 
\colhead{} & \colhead{} & \colhead{} & \colhead{(J2000) } & \colhead{ (J2000)} & \colhead{Position Angle} 
}
\startdata 
2 & 2005 Nov 12 & S255-IRS1 & 06\ 12\ 53.85 & +17\ 59\ 23.8 & 59.57 \\
3 & 2005 Dec 17 & S255-IRS1 & 06\ 12\ 53.85 & +17\ 59\ 23.8 & 99.57 \\
5 & 2006 Apr 8 & NGC 6334-V & 17\ 19\ 57.35 & $-$35\ 57\ 50.5 & 54.02  \\
6 & 2006 May 7 & NGC 6334-V & 17\ 19\ 57.35 & $-$35\ 57\ 50.5 & 79.57 \\
11 & 2006 Apr 10 & Oph-N9 & 16\ 27\ 13.31 & $-$24\ 41\ 32.4 & 68.88 \\
\enddata 

\end{deluxetable} 


%% file: tab2.tex


\begin{deluxetable}{lcccccccc}
\tabletypesize{\scriptsize}
\tablewidth{0pt}
\tablenum{2}
\tablecaption{Stellar and Clump Polarization and Photometry Measurements of NGC 6334-V}
\tablehead{
&& 
\colhead{R.A. Offset} & \colhead{Dec. Offset} & \colhead{R.A.} & \colhead{Dec.} &
\colhead{P} & \colhead{$\theta$} & \colhead{Mag$_{2.0 \mu m}$} \\
\colhead{Object} & \colhead{2MASS} &
\colhead{(arcsec)} & \colhead{(arcsec)} & \colhead{(J2000.0)} & \colhead{(J2000.0)} &
\colhead{(\%)} & \colhead{(degrees)} & \colhead{(mag)}  
}


\startdata
1 & 17195742-3558025 & $-$2.95 &  $-$20.79  & 17\ 19\ 54.42 & $-$35\ 58\ 02.4 & \nodata & \nodata  &  15.55 \\ 
2 & \nodata & $-$5.37 &  $-$19.34  & 17\ 19\ 57.22 & $-$35\ 58\ 01.0 & \nodata & \nodata  &  20.27 \\ 
3 & \nodata & $-$3.32 &  $-$18.84  & 17\ 19\ 57.39 & $-$35\ 58\ 00.5 & \nodata & \nodata & 19.27 \\ 
4 & \nodata & $-$6.15 &  $-$8.26  & 17\ 19\ 57.15 & $-$35\ 57\ 49.9 & \nodata & \nodata  &  20.35 \\ 
5 & \nodata &  \phs 2.17 &  $-$5.33  & 17\ 19\ 57.84 & $-$35\ 57\ 47.0 & \nodata & \nodata  &  20.21 \\ 
6 & \nodata &  \phs 2.50 &  $-$3.49  & 17\ 19\ 57.87 & $-$35\ 57\ 45.1 & $7 \pm 2$ & $144 \pm 10$  &  17.93 \\ 
7\tablenotemark{a} & \nodata &  \phs 5.23 &  $-$3.17  & 17\ 19\ 58.09 & $-$35\ 57\ 44.8 & $2 \pm 1$ & $83 \pm 3$  &  14.62 \\ 
8 & \nodata &  $-$13.24 &  $-$2.62  & 17\ 19\ 57.57 & $-$35\ 57\ 44.3 & \nodata & \nodata  &  16.92 \\ 
9 & \nodata &  $-$8.04 &  $-$2.47  & 17\ 19\ 57.00 & $-$35\ 57\ 44.1 & \nodata & \nodata &  19.96 \\ 
10 & \nodata &  $-$7.29 &  $-$2.23  & 17\ 19\ 57.06 & $-$35\ 57\ 43.9 & \nodata & \nodata  &  18.48 \\ 
11 & 17195766-3557416 &  \phs 0.00 &  \phs 0.00  & 17\ 19\ 57.66 & $-$35\ 57\ 41.6 & $3 \pm 1$ & $54 \pm 3$ & 12.08 \\ 
12 & \nodata &  \phs 0.78 & \phs 2.02  & 17\ 19\ 57.73 & $-$35\ 57\ 37.6 & $4 \pm 1$ & $127 \pm 7$ &  15.26 \\ 
13 & \nodata & $-$1.33 &  \phs 3.66  & 17\ 19\ 57.55 & $-$35\ 57\ 38.0 & \nodata & \nodata  &  16.16 \\ 
14 & \nodata & $-$2.33 &  \phs 4.86  & 17\ 19\ 57.47 & $-$35\ 57\ 36.78 & \nodata & \nodata &  16.94 \\ 
V-1 & \nodata & 4.5 & $-$7.1 & 17\ 19\ 58.03 & $-$35\ 57\ 48.8 & $71 \pm 3$ & $157 \pm 1$ & \nodata \\
V-2 & \nodata & 2.1 & $-$8.8 & 17\ 19\ 57.83 & $-$35\ 57\ 50.4 & $85 \pm 4$ & $163 \pm 2$ &  \nodata \\
V-3 & \nodata & $-$2.0 & $-$8.3 & 17\ 19\ 57.50 & $-$35\ 57\ 49.9 & $73 \pm 1$& $127 \pm 3$ & \nodata \\
V-4 & \nodata & 6.4 & $-$8.2 & 17\ 19\ 58.19 & $-$35\ 57\ 49.9 & $55 \pm 1$ & $166 \pm 2$ & \nodata \\
\enddata
\tablenotetext{a}{Star7 is elongated, possibly a binary.}
\end{deluxetable}


%% file: tab3.tex
\begin{deluxetable}{lccccccccc}
\tabletypesize{\scriptsize}
\tablewidth{0pt}
\tablenum{3}
\tablecaption{Stellar Polarization and Photometry Measurements of S255 IRS1}
\tablehead{
&&& 
\colhead{R.A. Offset} & \colhead{Dec. Offset} & \colhead{R.A.} & \colhead{Dec.} &
\colhead{P} & \colhead{$\theta$} & \colhead{Mag$_{2.0 \mu m}$} \\
\colhead{Object} & \colhead{NIRS\tablenotemark{a}} & \colhead{2MASS} &
\colhead{(arcsec)} & \colhead{(arcsec)} & \colhead{(J2000.0)} & \colhead{(J2000.0)} &
\colhead{(\%)} & \colhead{(degrees)} & \colhead{(mag)}  
}


\startdata
1 & NIRS9 & \nodata & \phs 4.49 & $-$10.74 & 06\ 12\ 54.33 & +17\ 59\ 12.4 & \nodata & \nodata & 14.70 \\ 
2 &  \nodata & \nodata &  \phs 4.22 &  $-$10.19  & 06\ 12\ 54.32 & +17\ 59\ 12.9 & \nodata & \nodata  &  16.89 \\ 
3\tablenotemark{b} & NIRS11 & 06125388+1759138 & $-$2.81 & $-$9.49 &  06\ 12\ 53.82  &  +17\ 59\ 13.6 &  $10 \pm 1$ &  $150 \pm 1$ & 14.29 \\ 
4 &  \nodata & \nodata & $-$5.63 & $-$8.96 &  06\ 12\ 53.63 & +17\ 59\ 14.1  & \nodata &  \nodata &  19.50 \\ 
5 & NIRS54 & \nodata &  \phs 5.23 & $-$8.89 &  06\ 12\ 54.39  &  +17\ 59\ 14.2  &  $8 \pm 1$ &  $148 \pm 3$  &  16.17 \\ 
6 &  \nodata & \nodata & $-$3.96 & $-$7.28 &  06\ 12\ 53.74 & +17\ 59\ 15.8  & $12 \pm 1$ &  $133 \pm 5$ &  17.04 \\ 
7 &  \nodata & \nodata & $-$7.30  &  $-$7.23  & 06\ 12\ 53.51  &  +17\ 59\ 15.9 &  $20 \pm 3$ &  $129 \pm 4$  & 18.16  \\
8 &  \nodata & \nodata &  $-$11.88 & $-$6.48 &  06\ 12\ 53.19 & +17\ 59\ 16.6  & \nodata & \nodata &  18.45  \\
9 & NIRS57 & \nodata & $-$2.53  &  $-$6.11  & 06\ 12\ 53.84  &  +17\ 59\ 17.0 & \nodata & \nodata &  14.68 \\ 
10 &  \nodata & \nodata &  \phs 5.41 & $-$6.01 &  06\ 12\ 54.40 & +17\ 59\ 17.1 & \nodata & \nodata & 18.06 \\ 
11 &  \nodata & \nodata & $-$5.11  &  $-$5.76 &  06\ 12\ 53.66  &  +17\ 59\ 17.3 &  $27 \pm 5$ &  $153 \pm 3$ &  18.90 \\ 
12 &  NIRS53 & \nodata &  \phs 5.55 & $-$5.54  & 06\ 12\ 54.41 & +17\ 59\ 17.6 & \nodata & \nodata & 15.47  \\
13 &  \nodata & \nodata & $-$7.50 & $-$5.39  & 06\ 12\ 53.49 & +17\ 59\ 17.7 &  $18 \pm 5$ &  $131 \pm 6$ & 18.48  \\
14 &  NIRS52 & \nodata & $-$3.83 & $-$5.36  & 06\ 12\ 53.75 & +17\ 59\ 17.7  &  $5 \pm 1$ &  $143 \pm 5$ &  14.48  \\
15 &  \nodata & \nodata &  $-$0.97  &  $-$4.41 &  06\ 12\ 53.95  &  +17\ 59\ 18.7  &  $6 \pm 1$ &  $143 \pm 4$  & 16.24  \\
16 &  \nodata & \nodata & \phs 5.63 & $-$3.48 &  06\ 12\ 54.41 & +17\ 59\ 19.6  & \nodata & \nodata &  19.33  \\
17 & \nodata & \nodata &  $-$5.12 & $-$3.15 &  06\ 12\ 53.66 & +17\ 59\ 19.9 & $5 \pm 1$ &  $154 \pm 4$ &  16.41  \\
18 &  \nodata & \nodata & $-$4.33  &  $-$2.92  & 06\ 12\ 53.72  &  +17\ 59\ 20.2 &  $10 \pm 1$ &  $176 \pm 4$ &  16.96  \\
19 &  \nodata & \nodata &  $-$9.33  &  $-$2.17  & 06\ 12\ 53.37  &  +17\ 59\ 20.9 &  \nodata & \nodata &  18.22  \\
20 &  \nodata & \nodata &  $-$1.02 & $-$2.07 &  06\ 12\ 53.95 & +17\ 59\ 21.0 &  $15 \pm 2$  &  $94 \pm 3$  & 17.02  \\
21 &  NIRS17 &  06125330+1759215 & $-$10.20 & $-$1.46 &  06\ 12\ 53.31 & +17\ 59\ 21.6  &  $3 \pm 1$ &  $158 \pm 2$ &  13.70 \\ 
22 &  \nodata & \nodata & \phs 1.06 & $-$0.46 &  06\ 12\ 54.09  &  +17\ 59\ 22.6 & \nodata & \nodata & 17.76  \\
23 &  \nodata & \nodata &  \phs 1.87 & $-$0.26 &  06\ 12\ 54.15  &  +17\ 59\ 22.8  & \nodata & \nodata & 16.84  \\
24 &  NIRS3 & \nodata &  \phs 0.00 &  \phs 0.00  & 06\ 12\ 54.02 & +17\ 59\ 23.1  &  $6 \pm 1$  &  $27 \pm 2$  & 13.57  \\
25 &  \nodata & \nodata & \phs 2.98 & \phs 0.56 & 06\ 12\ 54.23 & +17\ 59\ 23.7  & \nodata & \nodata &  19.64 \\ 
26 &  NIRS1 & 06125385+1759242 &  $-$2.40 &  \phs 0.62 &  06\ 12\ 53.85 & +17\ 59\ 23.7 &  $24 \pm 2$ &  $113 \pm 1$ &  12.74 \\ 
27 &  \nodata & \nodata & \phs 5.02 &  \phs 1.80 &  06\ 12\ 54.37 & +17\ 59\ 24.9 & \nodata & \nodata &  14.50  \\
28 &  \nodata & \nodata & \phs 4.18  & \phs 2.02 &  06\ 12\ 54.31  &  +17\ 59\ 25.1 & \nodata &  \nodata & 15.33  \\
29 &  \nodata & \nodata &  $-$8.14 &  \phs 2.09  & 06\ 12\ 53.45 & +17\ 59\ 25.2 & \nodata & \nodata  &  18.48  \\
30 &  \nodata & \nodata & \phs 1.58  & \phs 2.26 &  06\ 12\ 54.13  &  +17\ 59\ 25.4 & \nodata & \nodata & 17.49  \\
31 &  \nodata & \nodata & \phs 4.69 &  \phs 3.22 &  06\ 12\ 54.35  &  +17\ 59\ 26.3 & \nodata & \nodata & 16.62 \\ 
32 &  \nodata & \nodata &  $-$0.68  & \phs 3.74  & 06\ 12\ 53.97 & +17\ 59\ 26.8  & \nodata & \nodata &  17.23 \\ 
33 &  \nodata & \nodata &  $-$2.23 & \phs 3.98 &  06\ 12\ 53.86  &  +17\ 59\ 27.1 & \nodata & \nodata &  17.38 \\ 
34 &  \nodata & \nodata &  $-$9.89 &  \phs 4.04 &  06\ 12\ 53.33 & +17\ 59\ 27.1  & $10 \pm 2$ & $153 \pm 7$  & 18.20 \\ 
35 &  NIRS34 & \nodata & \phs 4.64 &  \phs 4.94 &  06\ 12\ 54.35  &  +17\ 59\ 28.0 & \nodata & \nodata & 14.74  \\
36 &  NIRS22 & 06125362+1759279 & $-$5.86 &  \phs 5.05 &  06\ 12\ 53.61 & +17\ 59\ 28.1 & $4 \pm 1$ &  $146 \pm 4$ &  13.66 \\ 
37 &  \nodata & \nodata & \phs 1.62 &  \phs 5.59 &  06\ 12\ 54.13 & +17\ 59\ 28.7 & $5 \pm 1$  & \nodata & 16.24 \\ 
38 &  \nodata & \nodata & \phs 5.28 & \phs 5.81 &  06\ 12\ 54.39  &  +17\ 59\ 28.9 & \nodata & \nodata & 16.30 \\ 
39 &  NIRS23 & 06125414+1759287 &  \phs 1.75 &  \phs 5.94 &  06\ 12\ 54.14 & +17\ 59\ 29.0 & $1 \pm 1$ & \nodata & 13.64 \\ 
40 &  \nodata & \nodata &  $-$2.87 &  \phs 5.98  & 06\ 12\ 53.82  &  +17\ 59\ 29.1  &  $5 \pm 1$ &  $174 \pm 5$ & 16.18 \\ 
41 & \nodata & \nodata &  $-$9.04 &  \phs 6.78 &  06\ 12\ 53.39  &  +17\ 59\ 29.9 & $7 \pm 1$ &  $133 \pm 4$ &  16.53 \\ 
42 &  NIRS24 & 06125394+1759303 & $-$0.97 & \phs 7.48 &  06\ 12\ 53.95 & +17\ 59\ 30.6 & $1 \pm 1$ &  $176 \pm 3$ &  13.39 \\ 
43 &  \nodata & \nodata &  \phs 5.86 &  \phs 7.57 &  06\ 12\ 54.43 & +17\ 59\ 30.7  &  $5 \pm 1$ & $78 \pm 1$ & 15.51 \\ 
44 & \nodata & \nodata &  $-$6.55  & \phs 7.63  & 06\ 12\ 53.56  &  +17\ 59\ 30.7  &  $3 \pm 1$  &  $96 \pm 9$ &  15.36 \\ 
45 & \nodata & \nodata & $-$4.35 &  \phs 9.28  & 06\ 12\ 53.72  &  +17\ 59\ 32.4 & $2 \pm 1$ & \nodata & 15.29 \\ 
46 & \nodata & \nodata &  \phs 4.76  & \phs 9.54 &  06\ 12\ 54.35 & +17\ 59\ 32.6 & \nodata & \nodata  &  18.71 \\ 
47 &  NIRS28 & 06125414+1759344 &  \phs 1.76 & \phs 11.52 &  06\ 12\ 54.14 & +17\ 59\ 34.6 & $4 \pm 1$ & $88 \pm 1$ & 12.84 \\ 

\enddata
\tablenotetext{a}{Tamura et al. (1991); Miralles et al. (1997); Itoh et al. (2001).}
\tablenotetext{b}{Nebulous object, probably a YSO.}

\end{deluxetable}

%% file: tab4.tex


\begin{deluxetable}{cccccccccc} 
\tabletypesize{\scriptsize}
\tablewidth{0pt} 
\tablenum{4}
\tablecaption{Parameters from Fits\tablenotemark{a}}
\tablehead{ 
& \colhead{$T_1$} & \colhead{$T_2$} &  & \colhead{$A_V\tablenotemark{b}$} & \colhead{Luminosity} & \colhead{Mass\tablenotemark{c}} & \colhead{$T_{\rm eff}$\tablenotemark{c}} & \colhead{$Q_0$\tablenotemark{c}} & \colhead{ZAMS} \\
\colhead{Star} & \colhead{(K)} & \colhead{(K)} & \colhead{$\tau_{9.6 \mu m}$} & \colhead{(mag)} & \colhead{(L$_\odot$)} & \colhead{(M$_\odot$)} & \colhead{(1000 K)} & \colhead{(s$^{-1}$)} &  \colhead{Sp. Type} 
}
\startdata 
NIRS1 & 326 & 993 & 1.55 & 20.2 & $1.4 \times 10^3$ & 6 -- 8 & 17 -- 20 & \nodata & B3 \\
NIRS3 & 381 & 972 & 3.53 & 46.1 & $2.4 \times 10^4$ & 14 -- 16 & 26 -- 32 & $1.0 \times 10^{46}$  -- $1.9 \times 10^{47}$ & B1 \\
\enddata 
\tablenotetext{a}{The least-squares fits for each YSO consist of two graybodies at different temperatures times an extinction curve. See the text for the other parameters.}
\tablenotetext{b}{The extinction $A_{2.0 \mu m}$ is $\sim 0.13 A_V$.}
\tablenotetext{c}{The first number in the cell is for the maximum luminosity on the radiative track and the second number is for the zero age main sequence (ZAMS).}

\end{deluxetable} 
